\documentclass[superscriptaddress,aps,prd,nofootinbib,reprint,preprintnumbers]{revtex4-1}
\pdfoutput=1

\usepackage{graphicx}
\usepackage[colorlinks,allcolors=blue]{hyperref} 
\usepackage{amsmath,amssymb,bm}
\usepackage[T1]{fontenc}
\usepackage{mathtools}
\usepackage[dvipsnames]{xcolor}
\begin{document}

\preprint{TU-1263}

%%%%%%%%%%%%%%% Title %%%%%%%%%%%%%%%%%%
\title{
Why \texorpdfstring{$w \ne -1$}{}? Anthropic Selection in a \texorpdfstring{$\Lambda$}{} + Axion Dark Energy Model
}
%%%%%%%%%%%%%%%%%%%%%%%%%%%%%%%%%%%%%%%%

%%%%%%%%%%%%%% Author %%%%%%%%%%%%%%%%%%
\author{Kai Murai}
\email{kai.murai.e2@tohoku.ac.jp}
\affiliation{Department of Physics, Tohoku University, Sendai, Miyagi 980-8578, Japan}
\author{Fuminobu Takahashi}
\email{fumi@tohoku.ac.jp}
\affiliation{Department of Physics, Tohoku University, Sendai, Miyagi 980-8578, Japan}
%%%%%%%%%%%%%%%%%%%%%%%%%%%%%%%%%%%%%%%

%%%%%%%%%%%%% Abstract %%%%%%%%%%%%%%%%
\begin{abstract}
We study a dark energy model composed of a bare negative cosmological constant and a single ultra-light axion, motivated by the string axiverse. 
Assuming that intelligent observers arise and observe, as in our universe, the onset of dark-energy-driven acceleration following matter domination, and that this acceleration persists to the present, we derive nontrivial constraints on both the axion mass and the bare cosmological constant. The axion mass is bounded from above to avoid fine-tuning of the initial misalignment angle near the hilltop, and from below because too light axions cannot achieve accelerated expansion due to their limited energy budget.
As a result, the anthropically allowed axion mass range typically lies around $m = \mathcal{O}(10)\, H_0$ for a decay constant close to the Planck scale, where $H_0$ is the observed value of the Hubble constant. In this framework, the dark energy equation-of-state parameter $w_0$ generically deviates from $-1$ by $\mathcal{O}(0.1)$, providing a natural explanation for why $w \ne -1$ may be expected. 
We also find that, for a decay constant slightly smaller than the Planck scale, the peak value of dark energy density is significantly smaller than the anthropic bound on the cosmological constant and can be close to the observed value.
These outcomes are intriguingly consistent with recent DESI hints of time-varying dark energy, and offer a compelling anthropic explanation within the $\Lambda$ + axion framework. 
\end{abstract}

%%%%%%%%%%%%%%%%%%%%%%%%%%%%%%%%%%%%%%%

\maketitle

%%%%%%%%%%%%%%% Section %%%%%%%%%%%%%%%
\section{Introduction}
%%%%%%%%%%%%%%%%%%%%%%%%%%%%%%%%%%%%%%%
The accelerated expansion of the universe is now well established through various cosmological observations, including Type Ia supernovae~\cite{SupernovaCosmologyProject:1998vns,SupernovaSearchTeam:1998fmf}, the cosmic microwave background (CMB)~\cite{Planck:2018vyg}, and large-scale structure~\cite{BOSS:2016wmc}. This phenomenon is commonly attributed to dark energy, which constitutes roughly 70\% of the current energy density of the universe. Despite its dominant role, the physical origin of dark energy remains one of the most profound mysteries in modern cosmology. A wide variety of theoretical models have been proposed to explain it, ranging from modifications of gravity~\cite{Carroll:2003wy} to dynamical scalar fields~\cite{Ratra:1987rm,Caldwell:1997ii}. Nevertheless, the simplest and most conservative candidate is a cosmological constant~\cite{Weinberg:1988cp}.

However, the cosmological constant faces a severe theoretical challenge. Naively, quantum field theory predicts a vacuum energy density of order $M_{\mathrm{Pl}}^4$, where $M_{\mathrm{Pl}} \simeq 2.4 \times 10^{18}$\,GeV is the reduced Planck mass, while observations suggest a value smaller by about 120 orders of magnitude. This enormous discrepancy constitutes the so-called cosmological constant problem~\cite{Weinberg:1988cp}. In addition, string theory suggests that constructing metastable vacua with positive vacuum energy is nontrivial, while vacua with negative cosmological constant are far more generic~\cite{Kachru:2003aw,Bousso:2000xa,Obied:2018sgi,Danielsson:2018ztv,Demirtas:2021ote}. Therefore, understanding the mechanism by which positive vacuum energy is uplifted from more generic negative-energy vacua remains a key challenge in connecting string theory to cosmological observations.

Among the proposed solutions to the cosmological constant problem, the anthropic principle~\cite{Weinberg:1987dv,Weinberg:1988cp} has gained significant attention, especially in the context of the string landscape~\cite{Susskind:2003kw,Polchinski:2006gy,Ellis:2009gx}. If the landscape contains a vast number of vacua with different vacuum energies, it is natural to expect that observers can only exist in regions of the multiverse where the cosmological constant falls within a narrow anthropic window: too large a value would suppress galaxy formation, while a negative value would lead to rapid recollapse. This line of reasoning offers a statistical explanation for the observed smallness of the cosmological constant.%
\footnote{
In fact, the observed amount of dark energy is smaller than the anthropic upper bound by about three orders of magnitude, which may call for an explanation~\cite{Sorini:2024tto}.
See Appendix A for a possible resolution within our framework.
}

In this work, we go beyond the single-component picture in the anthropic argument and consider a scenario in which dark energy arises from the sum of a bare negative cosmological constant and the energy of an ultra-light axion.%
\footnote{A related anthropic approach has been taken in the context of the strong CP problem, where the sum of a cosmological constant and an axion potential was considered as the effective cosmological constant~\cite{Takahashi:2008pu,Kaloper:2018kma}.}
Such ultra-light axions are generically predicted in string theory, where a large number of axions with masses spanning many orders of magnitude are expected—a scenario often referred to as the axiverse~\cite{Arvanitaki:2009fg}. If the axion is sufficiently light, it can remain displaced from its potential minimum over cosmological timescales and contribute to the present-day dark energy density~\cite{Svrcek:2006hf}.
The explanation of dark energy with a negative cosmological constant and an ultra-light axion has been explored in Refs.~\cite{Ruchika:2020avj,Luu:2025fgw}.
See also Refs.~\cite{Cardenas:2002np,Dutta:2018vmq,Visinelli:2019qqu,Calderon:2020hoc,Sen:2021wld,Adil:2023ara,Menci:2024rbq} for related works.

Our goal is not to determine the full anthropically allowed parameter space of the $\Lambda$ + axion dark energy model, which would be highly nontrivial due to the many free parameters and the complexity of possible cosmological histories. Instead, we restrict our attention to universes resembling our own, where 
intelligent observers arise and observe that the universe began accelerating due to dark energy after matter domination, and that this acceleration still continues today.
Within this context, we investigate what can be said about the properties of the axion, the dark energy equation-of-state parameter, and the dark energy density.

Under this restricted anthropic condition,
the total dark energy is required to lie below an anthropic upper bound on the cosmological constant.
Although this condition does not directly constrain each component individually, we show that it nonetheless leads to nontrivial constraints on both the axion mass and the bare cosmological constant.
Specifically, the axion mass is bounded from above because large masses would require the initial misalignment angle to be finely tuned near the hilltop in order to avoid early relaxation and recollapse. Conversely, the axion mass is bounded from below because, for too small masses, the axion cannot achieve accelerated expansion due to its small energy budget.
These dual constraints result in a finite range of allowed axion masses and bare vacuum energies.

Remarkably, within this constrained parameter space, the dark energy equation-of-state parameter $w$ is generically predicted to deviate from $-1$ by $\mathcal{O}(0.1)$. This deviation arises naturally from the anthropic selection and is consistent with recent DESI results hinting at time-varying dark energy~\cite{DESI:2024mwx,DESI:2025zgx,DESI:2025zpo}, providing a compelling explanation within the $\Lambda$ + axion framework.
See Refs.~\cite{Tada:2024znt,Yin:2024hba,Cortes:2024lgw,Bhattacharya:2024hep,Mukherjee:2024ryz,Notari:2024rti,Jia:2024wix,Hernandez-Almada:2024ost,Bhattacharya:2024kxp,Berbig:2024aee,Wolf:2025jlc,Chakraborty:2025syu,Borghetto:2025jrk,Nakagawa:2025ejs,Lee:2025yvn,Colgain:2025nzf,Santos:2025wiv} for various explanations of the DESI results.

%%%%%%%%%%%%%%% Section %%%%%%%%%%%%%%%
\section{\texorpdfstring{$\Lambda$}{} + Axion dark energy model}
\label{sec: anthropic}
%%%%%%%%%%%%%%%%%%%%%%%%%%%%%%%%%%%%%%%

As a concrete realization of the anthropic scenario outlined in the introduction, we consider a model in which dark energy is composed of a negative cosmological constant, $\rho_\Lambda < 0$, and a single axion field $\phi$. 
Here, the presence of a negative cosmological constant is an assumption motivated by string theory~\cite{Kachru:2003aw,Bousso:2000xa,Obied:2018sgi,Danielsson:2018ztv,Demirtas:2021ote}, while $\rho_\Lambda < 0$ is preferred from the observational data in the framework of $\Lambda$ + axion dark energy~\cite{Luu:2025fgw}.
While the string axiverse generically contains multiple axions, we focus on a single axion field for simplicity. This approximation is justified if the axion mass spectrum is hierarchical: heavier axions would have already settled into their minima, while much lighter ones remain effectively frozen, and their energy densities can be effectively absorbed into $\rho_\Lambda$. 

The axion potential is taken to be 
\begin{align}
    V(\phi)
    =
    m^2 f^2 \left[ 1 + \cos \left( \frac{\phi}{f} \right) \right]
    \ ,
\end{align}
where $m$ is the axion mass, $f$ is its decay constant, and we choose the origin, $\phi = 0$, to be the maximum of the potential.
We set $f = M_\mathrm{Pl}$ throughout the main text, and we will consider the case with smaller decay constants in a later discussion. 
We also assume that the axion field is spatially homogeneous.

For our analysis, we treat
$\rho_\Lambda$,  $\theta_\mathrm{i} \equiv \phi_\mathrm{i}/f$, and $m$ as independent random variables with flat priors for $\rho_\Lambda$, $\theta_\mathrm{i}$, and $\ln m$, while fixing all other parameters to the observed values.
Here, $\phi_\mathrm{i}$ is the initial value of $\phi$.

In an expanding universe, the axion follows the equation of motion,
\begin{align}
    \ddot{\phi} + 3 H \dot{\phi} - m^2 f \sin \left(\frac{\phi}{f} \right)
    =
    0
    \ ,
\end{align}
where the dots denote  derivatives with respect to the cosmic time $t$, and $H$ is the Hubble parameter determined by the Friedmann equation,
\begin{align}
    3 M_\mathrm{Pl}^2 H^2
    =
    \rho_\mathrm{DE} + \rho_m
    \equiv 
    \rho_\Lambda + \rho_\phi +  \rho_m 
    \ .
\end{align}
Here, $\rho_m$ and $\rho_\phi$ are the energy densities of the non-relativistic matter and the axion, respectively.
We neglect the radiation component, which is subdominant during the epoch of interest.

To compare cosmic histories with different parameters, we define the present time $t_0$ by $\rho_m(t_0) = \rho_{m,0}$, where $\rho_{m,0} = 3 M_\mathrm{Pl}^2 H_0^2 \Omega_m$ is the observed matter density at present.
Here, $\Omega_m \simeq 0.3$ is the observed matter density parameter, and $H_0 \simeq 67\,\mathrm{km/s/Mpc}$ is the observed Hubble constant.
We normalize the scale factor as $a(t_0) = 1$ so that the matter density evolves as
\begin{align}
    \rho_m = \rho_{m,0}\, a^{-3} \ .
\end{align}
On the other hand, the axion energy density is given by 
\begin{align}
    \rho_\phi 
    =
    \frac{1}{2} \dot{\phi}^2 + V(\phi)
    \ .
\end{align}

Anthropic considerations for a conventional cosmological constant alone impose an upper bound on its energy density, $\rho_\Lambda < \rho_\mathrm{\Lambda}^{\rm max}$, derived from the requirement that galaxies can form~\cite{Weinberg:1987dv,Weinberg:1988cp}.
In the following, we adopt the upper bound from the star formation history, $\rho_\mathrm{\Lambda}^{\rm max} = 2306 \rho_\mathrm{DE,obs}$~\cite{Sorini:2024tto} with $\rho_\mathrm{DE,obs} = 3 M_\mathrm{Pl}^2 H_0^2 \Omega_\mathrm{DE}$ and $\Omega_\mathrm{DE} \simeq 0.7$.
A lower bound also follows from the requirement that 
the universe should not recollapse too soon after galaxy formation, allowing sufficient time for the development of intelligent life.

In this work on the axion dark energy, we further require that intelligent life observe accelerated expansion of the universe that persists to the present epoch.
The accelerated expansion is realized when 
\begin{align}
    w_\mathrm{tot} 
    \equiv 
    \frac{p_\mathrm{DE}}{\rho_m + \rho_\mathrm{DE}}
    < 
    - \frac{1}{3}
    \ ,
\end{align}
where $p_\mathrm{DE} = \dot{\phi}^2/2 - V(\phi) - \rho_\Lambda$ is the pressure of the dark energy component.
In our scenario, the total dark energy component is given by the sum of a negative cosmological constant and an ultra-light axion and it satisfies $p_\mathrm{DE} \geq - \rho_\mathrm{DE}$.
Thus, $\rho_\mathrm{DE}$ must become larger than $\rho_m/2$ at some point before the current time to realize $w_\mathrm{tot} < -1/3$.
As a result, we obtain the condition for $\rho_\mathrm{DE,i}$ to satisfy our requirement as
\begin{align} 
    \label{eq:window}
    \frac{\rho_{m,0}}{2} < \rho_\mathrm{DE,\mathrm{i}} \lesssim \rho_\mathrm{\Lambda}^{\rm max},
\end{align} 
where $\rho_\mathrm{DE,\mathrm{i}}$ is the initial energy density of dark energy.
We refer to this as a conditional anthropic bound.
The lower bound reflects the need for the axion to uplift the negative vacuum energy and initially behave like a cosmological constant in order to drive accelerated expansion. 
The upper bound ensures that galaxy formation occurs before the axion starts oscillating. 
Note that the upper bound, $\rho_\Lambda^\mathrm{max}$, is derived under the assumption that the dark energy component solely consists of a cosmological constant.
In our scenario, however, the axion should remain around the hilltop for accelerated expansion to persist until now, and so the dark energy density is expected to evolve only slowly.
Therefore, we adopt $\rho_\Lambda^\mathrm{max}$ as the effective upper bound on the initial value of the total dark energy density.
Since the axion remains frozen in the early universe, we can set the initial time arbitrarily as long as it is well before the axion starts to evolve.
Due to the upper bound on $\rho_\mathrm{DE,\mathrm{i}}$, the dark energy is negligible until well after the matter-radiation equality, and it does not affect big bang nucleosynthesis and recombination.

Since we fix $f = M_\mathrm{Pl}$, the Hubble parameter is always smaller than the axion mass $m$ during the dark-energy-dominated epoch.
In particular, if $m$ is much larger than the Hubble parameter, the axion would have rolled down the potential by now, and the universe would have ceased accelerating unless the initial field value were sufficiently close to the hilltop.
This degree of fine-tuning becomes increasingly severe for larger values of $|\rho_\Lambda|$, as a larger axion mass is required to uplift the total vacuum energy into the conditional anthropic window (\ref{eq:window}).

While we can evaluate the axion dynamics numerically, let us first obtain an approximate analytic description.
We assume that the axion remains near the hilltop even at the present time.
Then, $\rho_\mathrm{DE}$ is approximately given by
\begin{align}
    \rho_\mathrm{DE} 
    \simeq 
    \rho_\Lambda + 2 m^2 f^2
    \ .
\end{align}
With this approximation, we can solve the Friedmann equation to obtain
\begin{align}
    a(t)
    \simeq
    \left( \frac{\rho_{m,0}}{\rho_\mathrm{DE}} \right)^{1/3}
    \sinh^{2/3} \left( \frac{3}{2} H_\mathrm{DE} t \right)
    \ ,
\end{align}
%%
%where 
with
\begin{align}
    H_\mathrm{DE}
    \equiv
    \sqrt{\frac{\rho_\mathrm{DE}}{3 M_\mathrm{Pl}^2}}
    \ .
\end{align}
Then, the equation of motion becomes
\begin{align}
    \ddot{\phi} + 
    3 H_\mathrm{DE} \frac{\cosh\left(\frac{3}{2}H_\mathrm{DE}t\right)}{\sinh\left(\frac{3}{2}H_\mathrm{DE}t\right)}\dot{\phi}
    - m^2 \phi
    \simeq 0
    \ ,
\end{align}
where we used $\sin(\phi/f) \simeq \phi/f$ for $\phi \ll f$.
With the initial condition 
\begin{align}
    \phi(t=0) = \phi_\mathrm{i}
    \ , \quad 
    \dot{\phi}(t=0) = 0
    \ ,
\end{align}
the equation of motion can be solved to yield
\begin{align}
    \phi(t) 
    &=
    \phi_\mathrm{i} \frac{\sinh(\sqrt{1+A}\tilde{t})}{\sqrt{1+A} \sinh(\tilde{t})}
    \ ,
    \label{eq: phi approx}
\end{align}
where
\begin{align}
    A \equiv \frac{4m^2}{9H_\mathrm{DE}^2}
    \ , \quad 
    \tilde{t} \equiv \frac{3}{2}H_\mathrm{DE}t
    \ .
\end{align}
Thus, at the present time, we obtain
\begin{align}
    \phi(t_0)
    =&
    \phi_\mathrm{i} \frac{\sinh(\sqrt{1+A}\tilde{t}_0)}{\sqrt{1+A} \sinh(\tilde{t}_0)}
    \ ,
    \\
    \dot{\phi}(t_0)
    =&
    \frac{3}{2} H_\mathrm{DE} \phi_\mathrm{i} 
    \nonumber \\
    & \times
    \left[ 
        \frac{\cosh(\sqrt{1+A}\tilde{t}_0)}{\sinh(\tilde{t}_0)}
        -
        \frac{\cosh(\tilde{t}_0)\cosh(\sqrt{1+A}\tilde{t}_0)}{\sqrt{1+A}\sinh^2(\tilde{t}_0)}
    \right]
    \ ,
\end{align}
with
\begin{align}
    \tilde{t}_0 
    \equiv 
    \frac{3}{2} H_\mathrm{DE} t_0
    =
    \mathrm{arcsinh}\left( \sqrt{\frac{\rho_\mathrm{DE}}{\rho_{m,0}}} \right)
    \ .
\end{align}
The dark energy equation-of-state parameter at the present time $t=t_0$ is given by
\begin{align}
    w_0
    &=
    \left. 
    \frac{\frac{1}{2}\dot{\phi}^2 - \rho_\Lambda - V(\phi)}{\frac{1}{2}\dot{\phi}^2 + \rho_\Lambda + V(\phi)}
    \right|_{t = t_0}
    \ .
    \label{eq: w0}
\end{align}

\begin{figure}[t!]
    \begin{center}  
        \includegraphics[width=70mm]{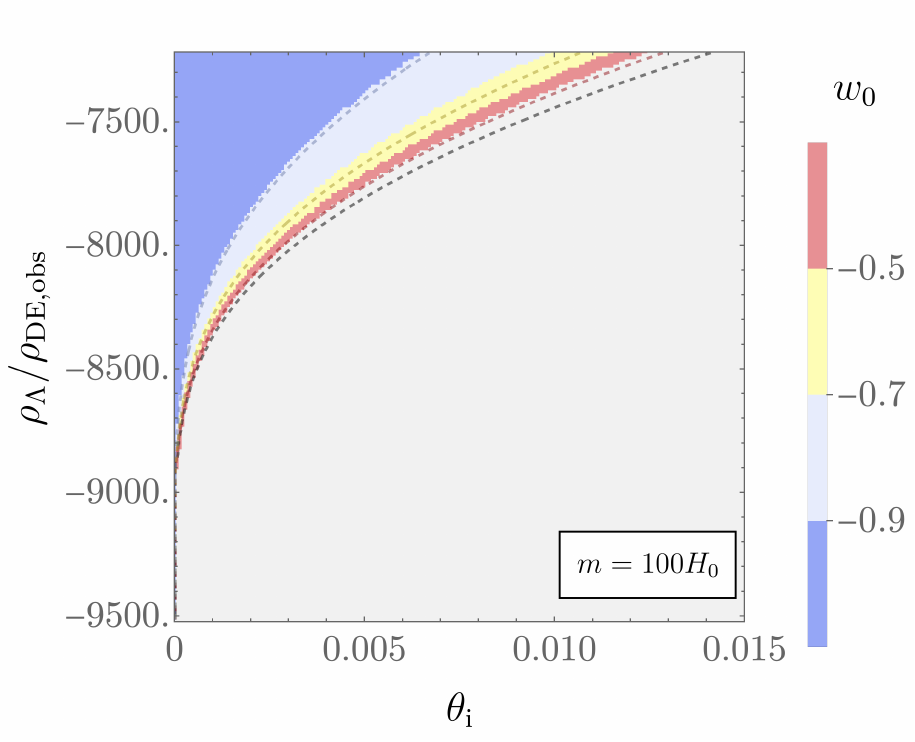}
        \\
        \vspace{2mm}
        \includegraphics[width=70mm]{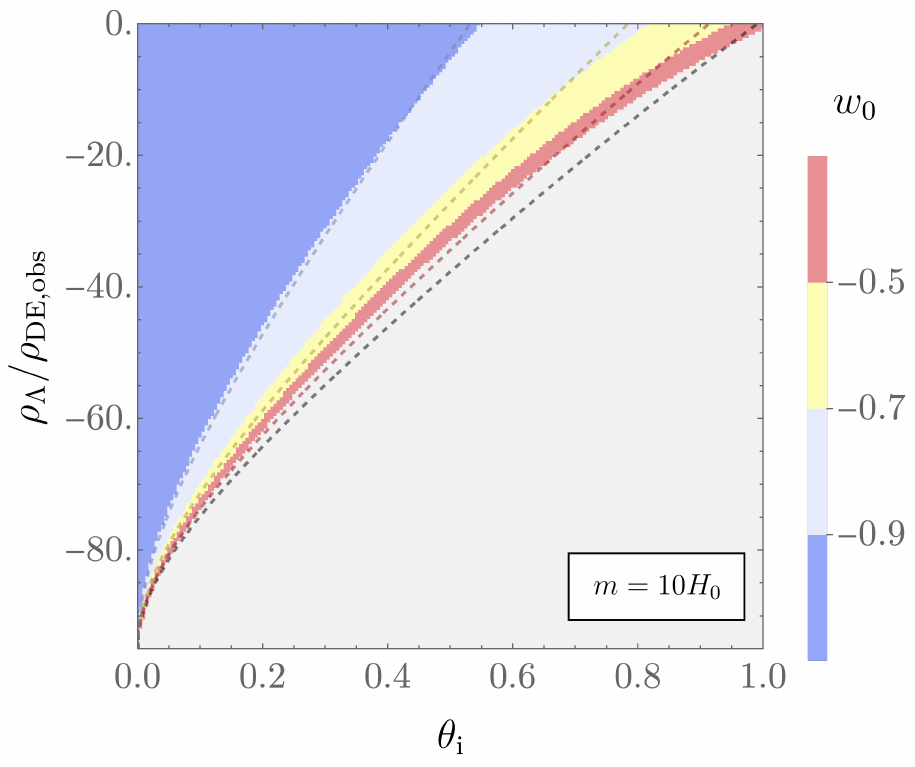}
        \\
        \vspace{2mm}
        \includegraphics[width=70mm]{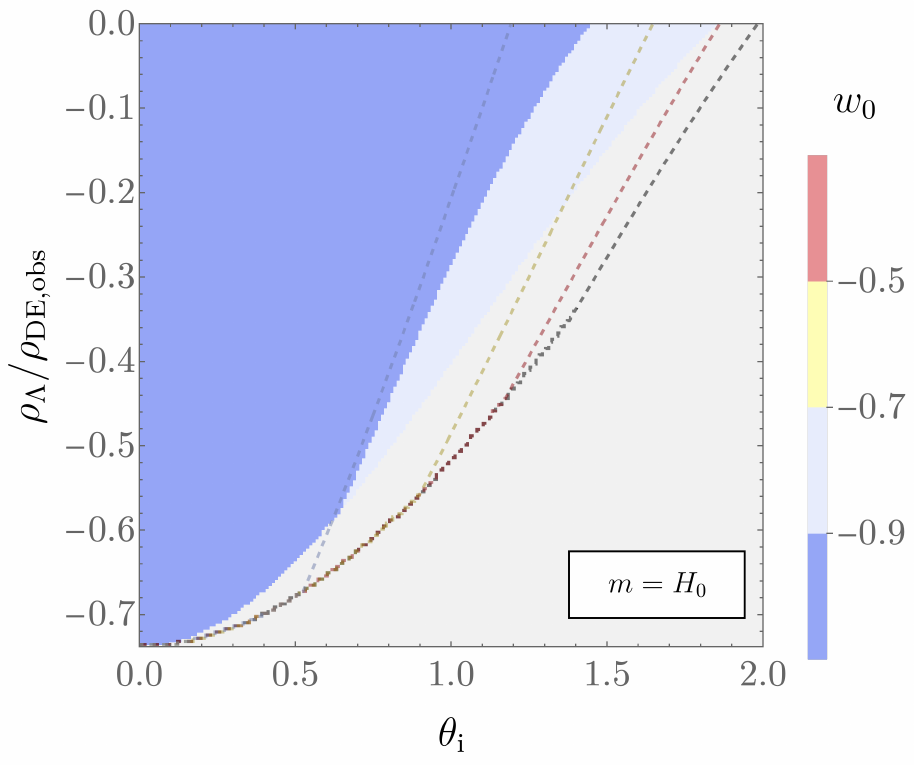}
        \end{center}
    \caption{%
        Colored regions indicate the predicted values of $w_0$, and dashed lines show analytic estimates of the boundaries between them. The axion mass is taken to be $m = 100 H_0$, $10 H_0$, and $H_0$ from top to bottom. The gray-shaded region in each panel is excluded due to the requirement for the accelerated expansion to persist until the current time.
    }
    \label{fig: w} 
\end{figure}

We evaluate the evolution of the axion and $w_0$ both numerically and analytically.
In Fig.~\ref{fig: w}, we show the parameter region allowed by the conditional anthropic bound for $m = 100 H_0$, $10 H_0$, and $H_0$. 
We fix the range of $\rho_\Lambda$ following the condition~\eqref{eq:window}.
Note that the scales of the horizontal and vertical axes are different in the three panels.
In the gray-shaded regions, our requirements are not satisfied: the universe never experiences accelerated expansion, or the accelerated expansion does not last 
until the current time.
We also show $w_0$ by the color shading for the allowed regions.
Smaller values of $\theta_\mathrm{i}$ lead to smaller $w_0$, as the axion remains closer to the hilltop.
The required tuning of $\theta_\mathrm{i}$ becomes more severe for small $\rho_\Lambda$ (i.e., large $|\rho_\Lambda|$) because the Hubble parameter is small, which allows the axion to roll down the potential more easily.
For $m = H_0$, $w_0$ cannot deviate significantly from $-1$, namely, the axion should remain around the hilltop at the current time, since otherwise the accelerated universe cannot be achieved due to the small energy budget of the axion.
%%%%%%%%%%%%%%%%%% Figure %%%%%%%%%%%%%%%%%%
%%%%%%%%%%%%%%%%%%%%%%%%%%%%%%%%%%%%%%%%%%%%

We also show the results of the analytic estimate with dashed lines.
Below the black dashed lines, $\rho_\mathrm{DE,i} < \rho_{m,0}/2$ or $w_0 > -1/3$ holds.
The colored dashed lines correspond to $w_0 = -0.9, -0.7, -0.5$ from left to right.
For $m = H_0$, the black dashed line overlaps with the colored lines.
While the analytic estimate of $w_0$ agrees well with the numerical results for small $w_0$ and large $m$, it worsens in the opposite case, where the approximations of $\phi \ll f$ and $\rho_\phi = \mathrm{const}.$ break down.

\begin{figure}[t!]
    \begin{center}  
        \includegraphics[width=80mm]{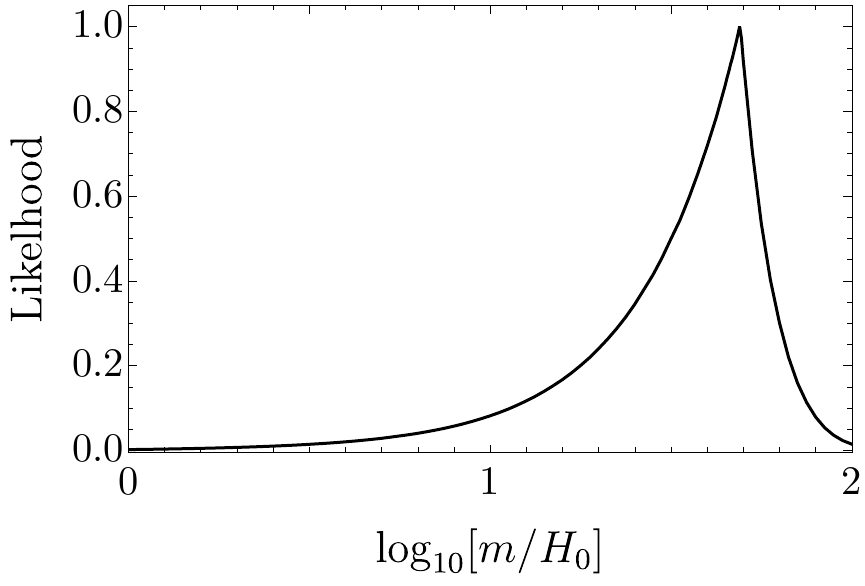}
        \end{center}
    \caption{%
        Likelihood of $\ln m$ marginalized over $\theta_\mathrm{i}$ and $\rho_\Lambda$.
        The vertical axis is normalized by the maximum value.
        Lower masses are disfavored due to the limited range of $\rho_\Lambda$, while higher masses are suppressed due to fine tuning of $\theta_\mathrm{i}$.
    }
    \label{fig: mass likelihood} 
\end{figure}

With the flat priors for $\theta_\mathrm{i}$ and $\rho_\Lambda$, we can evaluate the likelihood of $\ln m$ by integrating the area of the colored regions in Fig.~\ref{fig: w} over a denser sampling of the mass.
We show the likelihood of $\ln m$ in Fig.~\ref{fig: mass likelihood}.
The vertical axis is normalized so that the maximum value equals unity.
We find a peak at $m = m_\mathrm{peak} \simeq 49.2 H_0$, which is determined by $2 m^2 f^2 = \rho_\Lambda^\mathrm{max}$.
This behavior can be understood as follows. 
Assuming that the axion stays near the hilltop, 
the allowed range of $\rho_\Lambda$ is limited as $- 2 m^2 f^2 + \rho_{m,0}/2 \lesssim \rho_\Lambda \lesssim 
{\rm Min}[0,\rho_{\Lambda}^{\rm max}-2 m^2 f^2]$.
For small $m$ satisfying $\rho_{m,0}/2 \ll 2 m^2 f^2 < \rho_{\Lambda}^{\rm max}$, the interval of $\rho_\Lambda$ is approximately proportional to $m^2$.
So, the likelihood increases roughly $\propto m^2$.
For large $m$ satisfying $2 m^2 f^2 > \rho_{\Lambda}^{\rm max}$, while the interval of the allowed $\rho_\Lambda$ becomes approximately equal to $\rho_\mathrm{\Lambda}^\mathrm{max}$, independent of $m$, the required tuning of $\theta_\mathrm{i}$ near the hilltop becomes exponentially severe, suppressing the likelihood.
Note that the peak position of the axion mass distribution depends on $\rho_\Lambda^\mathrm{max}$ as $m_\mathrm{peak}^2 = \rho_\Lambda^\mathrm{max}/(2 f^2)$.
In particular, if there exists some unknown anthropic condition that imposes a much tighter upper bound on $\rho_\Lambda$, which is close to the observed dark energy density, the peak of likelihood comes to $m_\mathrm{peak} = \mathcal{O}(H_0)$.
Moreover, this estimate for $m_\mathrm{peak}$ is further corrected for smaller $f$.
See Appendix~\ref{app: parameters} for the different choices of $\rho_\Lambda^\mathrm{max}$ and $f$.
%%%%%%%%%%%%%%%%%% Figure %%%%%%%%%%%%%%%%%%
%%%%%%%%%%%%%%%%%%%%%%%%%%%%%%%%%%%%%%%%%%%%

\begin{figure}[t!]
    \begin{center}  
        \includegraphics[width=80mm]{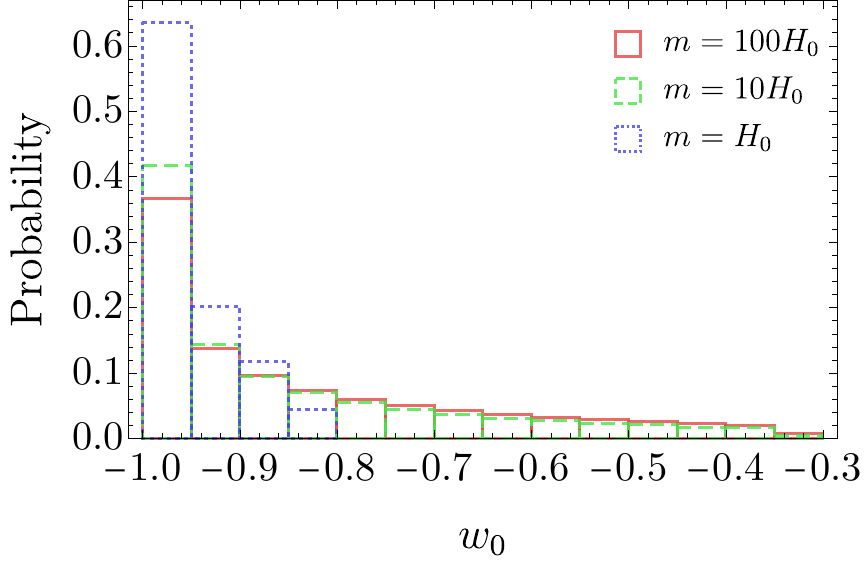}
        \end{center}
    \caption{%
       Histograms of $w_0$ for the three axion masses, $m = 100 H_0$, $10 H_0$, and $H_0$, are shown in different colors.
    }
    \label{fig: histogram} 
\end{figure}

In Fig.~\ref{fig: histogram}, we show the probability distribution of $w_0$ for fixed $m$.
Here, we sample $w_0$ from $-1$ to $-0.3$ in increments of $0.05$. 
Note that the accelerated expansion at the current time implies $w_0 < -1/3$.
We evaluate the probability by integrating the area with corresponding values of $w_0$ in Fig.~\ref{fig: w} and normalize it such that the total equals unity.
We find that while the most likely value of $w_0$ lies close to $-1$, the deviation by ${\cal O}(0.1)$ is statistically significant especially for $m = 100H_0$ and $10H_0$. 
Given that the mass likelihood of $\ln m$ has a peak at $10H_0 < m < 100H_0$, our results suggest that, in a universe where intelligent observers can observe the accelerated expansion, the dark energy equation-of-state parameter is likely to deviate from $-1$ at the level of ${\cal O}(0.1)$.
%%%%%%%%%%%%%%%%%% Figure %%%%%%%%%%%%%%%%%%
%%%%%%%%%%%%%%%%%%%%%%%%%%%%%%%%%%%%%%%%%%%%

\begin{figure}[t!]
    \begin{center}  
        \includegraphics[width=80mm]{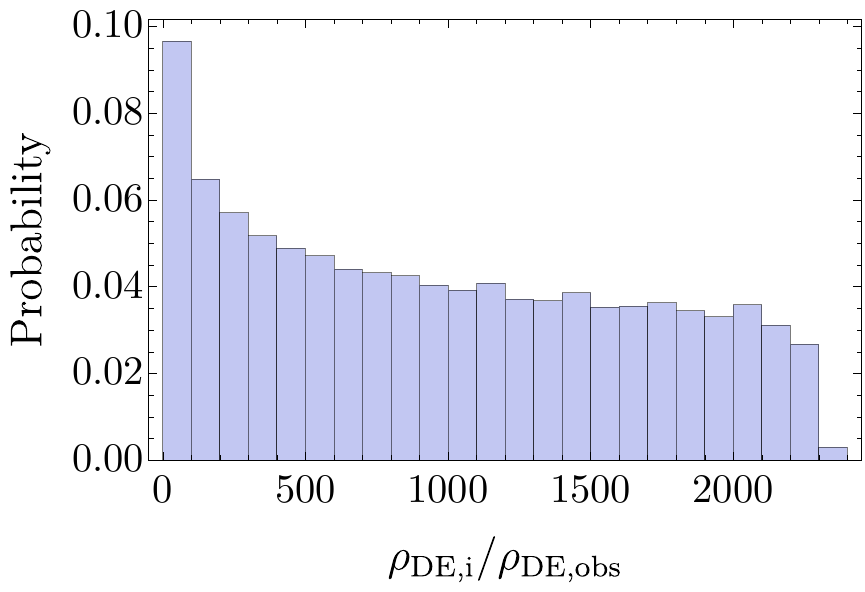}
        \end{center}
    \caption{%
       Histogram of $\rho_\mathrm{DE,i}$ marginalized over $\theta_\mathrm{i}$, $\rho_\Lambda$, and $\ln m$.
    }
    \label{fig: rhoDE histogram} 
\end{figure}

Finally, we show the probability distribution of $\rho_\mathrm{DE,i}$ with the flat priors for $\theta_\mathrm{i}$, $\rho_\Lambda$, and $\ln m$ in Fig.~\ref{fig: rhoDE histogram}.
In a similar way to Fig.~\ref{fig: histogram}, we integrate the area of the colored regions with $\rho_\mathrm{DE,i}$ in each bin for different values of $\ln m$ and sum the area over $\ln m$ sampled with a uniform interval.
Here, we take the bin for $0 < \rho_\mathrm{DE,i}/\rho_\mathrm{DE,obs} < 2400$ in increment of $100$.
Note that the rightmost bin corresponds to $2300 < \rho_\mathrm{DE,i}/\rho_\mathrm{DE,obs} < 2306$ due to the requirement~\eqref{eq:window} and thus has a lower probability.
The probability distribution for $\rho_\mathrm{DE,i}$ spreads out over the region allowed by Eq.~\eqref{eq:window}, and thus we cannot account for the hierarchy between $\rho_\mathrm{DE,obs}$ and $\rho_\Lambda^\mathrm{max}$ in this case.
On the other hand, as we discuss in Appendix~\ref{app: parameters}, $\rho_\mathrm{DE,i}$ close to the observed value is favored when we choose a smaller $f$ such as $f = 0.1 M_\mathrm{Pl}$.
Thus, for $f= {\cal O}(0.1) M_\mathrm{Pl}$,  $w_0$ naturally deviates from $-1$ by ${\cal O}(0.1)$, and the dark energy density approaches the observed value,  resolving the hierarchy between $\rho_{\rm DE,0}$ and $\rho_{\Lambda}^{\rm max}$. 
%%%%%%%%%%%%%%%%%% Figure %%%%%%%%%%%%%%%%%%
%%%%%%%%%%%%%%%%%%%%%%%%%%%%%%%%%%%%%%%%%%%%

%%%%%%%%%%%%%%% Section %%%%%%%%%%%%%%%
\section{Discussion and Conclusions}
\label{sec:conclusions}
%%%%%%%%%%%%%%%%%%%%%%%%%%%%%%%%%%%%%%%

So far, we have focused on the case where the bare cosmological constant is restricted to be negative. This simplifying assumption is well-motivated in the context of string theory, where vacua with negative vacuum energy are favored. Nonetheless, it is natural to consider generalizations in which the bare cosmological constant can also take positive values.
We have also assumed that observers exist in the universe and observe cosmic acceleration. 
Relaxing this assumption is another possible extension.
If one allows an unrestricted range of positive values for the bare cosmological constant, or does not impose the condition that observers see an accelerating universe, the anthropic argument no longer provides a lower bound on $m$ when assuming a flat measure in $\ln m$ for the axion mass.
This is because the probability distribution remains constant for arbitrarily small $m$, leading to a divergent probability. This requires some form of regularization, either through a modified measure or by imposing a prior on the axion mass.
However, if the allowed positive values of the cosmological constant are limited to those smaller than, for instance, $\rho_{m,0}/2$, our analysis can be straightforwardly extended without encountering this issue. In such cases, the lower bound on $m$ still emerges from the conditional anthropic condition.

Alternatively, if we restrict ourselves to the case of a vanishing cosmological constant (i.e., the pure axion dark energy model), we obtain qualitatively similar results for the likelihood functions of the dark energy equation-of-state parameter, axion mass, and dark energy density.

It should also be noted that including both eternally expanding universes and those that eventually recollapse in the anthropic argument is itself nontrivial. 
Unlike the case of a positive cosmological constant, where galaxies can persist indefinitely once formed, a universe with negative vacuum energy inevitably undergoes recollapse within a finite time. 
Whether such recollapsing universes should be assigned the same anthropic weight as eternally expanding ones is unclear, and addressing this issue may require a more refined treatment of the multiverse measure.
A proper assessment of these issues would involve a careful reevaluation of the prior distributions over both the cosmological constant and the axion parameters. We leave such investigations for future work.

We have focused on a dark energy model composed of a cosmological constant and an axion. Axions are also known to play other roles in cosmology, such as contributing to dark matter and dark radiation, and anthropic arguments concerning their abundances have been discussed in Refs.~\cite{Hellerman:2005yi,Takahashi:2019ypv}. It would be interesting to extend our analysis to include these additional dark components.

In this paper, we have explored a dark energy model composed of a bare negative cosmological constant and a single ultra-light axion, motivated by the string axiverse.
Assuming that intelligent observers exist and observe an accelerating universe, a conditional anthropic bound~(\ref{eq:window}) naturally follows, which constrains the initial total dark energy density $\rho_\mathrm{DE,i}$ to lie within a range compatible with both galaxy formation and cosmic acceleration.
Under this condition, we derived nontrivial anthropic constraints on both the axion mass and the bare cosmological constant.
The axion mass is bounded from above to avoid fine-tuning of the initial misalignment angle near the hilltop, and from below to 
achieve accelerated expansion. 

We have found that, within the anthropically allowed region, the axion mass typically lies around ${\cal O}(10) H_0$ for a decay constant near the Planck scale, and it slightly decreases  for smaller decay constants (see Appendix \ref{app: parameters}).
While the most likely value of the dark energy equation-of-state parameter $w_0$ lies close to $-1$, the deviation by ${\cal O}(0.1)$ is statistically significant.
This provides a natural anthropic explanation for why $w_0 \ne -1$ may be expected, and is intriguingly consistent with recent DESI results hinting at time-varying dark energy~\cite{DESI:2024mwx,DESI:2025zgx,DESI:2025zpo}.
Our results thus offer a concrete realization of how the string theory landscape, together with a light axion, can shape the nature of dark energy in our universe.

%%%%%%%%%%% Acknowledgments %%%%%%%%%%%
\section*{Acknowledgments}
We thank Wen Yin for helpful discussion on the DESI results.
We also thank the anonymous referee for suggesting a comparison with the latest observational data.
This work is supported by JSPS Core-to-Core Program (grant number: JPJSCCA20200002) (F.T.), JSPS KAKENHI Grant Numbers 20H01894 (F.T.), 20H05851 (F.T.), 23KJ0088 (K.M.), 24K17039 (K.M.), and 25H02165 (F.T.).
This article is based upon work from COST Action COSMIC WISPers CA21106, supported by COST (European Cooperation in Science and Technology).
%%%%%%%%%%%%%%%%%%%%%%%%%%%%%%%%%%%%%%

\appendix

%%%%%%%%%%%%%%% Section %%%%%%%%%%%%%%%
\section{Dependence on the parameters}
\label{app: parameters}
%%%%%%%%%%%%%%%%%%%%%%%%%%%%%%%%%%%%%%%

Here, we discuss the dependence of our results on the parameters such as $\rho_\Lambda^\mathrm{max}$ and $f$.

%%%%%%%%%%%%%%% Section %%%%%%%%%%%%%%%
\subsection{Smaller \texorpdfstring{$\rho_\Lambda^\mathrm{max}$}{}}
% \label{}
%%%%%%%%%%%%%%%%%%%%%%%%%%%%%%%%%%%%%%%

\begin{figure}[t!]
    \begin{center}  
        \includegraphics[width=80mm]{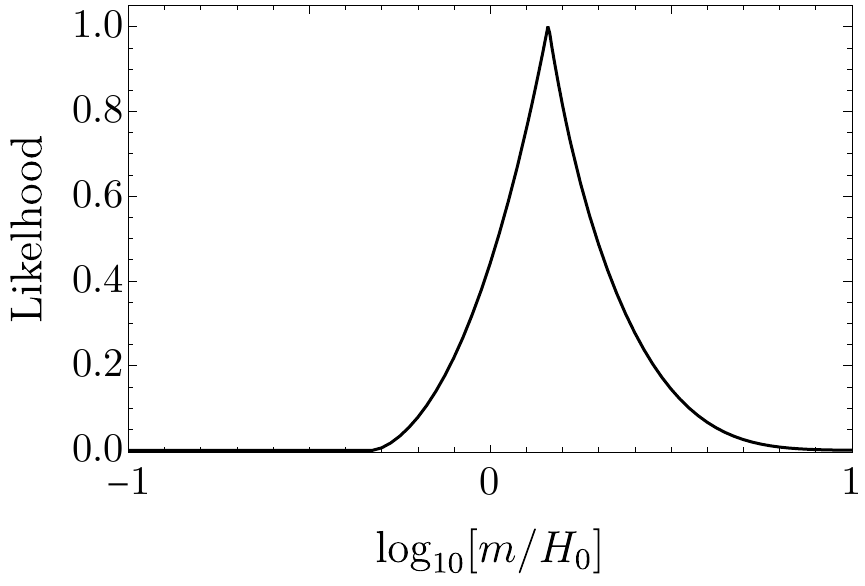}
        \end{center}
    \caption{%
        Same as Fig.~\ref{fig: mass likelihood} except for $\rho_\Lambda^\mathrm{max} = 2 \rho_\mathrm{DE,obs}$.
    }
    \label{fig: mass likelihood rhoLambda} 
\end{figure}

As discussed in the main text, the mass likelihood for $\ln m$ has a peak at $m = m_\mathrm{peak}$ determined by $2 m_\mathrm{peak}^2 f^2 = \rho_\Lambda^\mathrm{max}$.
Thus, if we could set a more stringent upper bound on $\rho_\Lambda$ from refined anthropic considerations or other reasoning, $m_\mathrm{peak}$ would become smaller.
To illustrate this expectation, we show the mass likelihood assuming a test value of $\rho_\Lambda^\mathrm{max} = 2 \rho_\mathrm{DE,obs}$ in Fig.~\ref{fig: mass likelihood rhoLambda}.
As expected, we obtain a peak at $m_\mathrm{peak} \simeq 1.4 H_0$.
%%%%%%%%%%%%%%%%%% Figure %%%%%%%%%%%%%%%%%%
%%%%%%%%%%%%%%%%%%%%%%%%%%%%%%%%%%%%%%%%%%%%

%%%%%%%%%%%%%%% Section %%%%%%%%%%%%%%%
\subsection{Smaller \texorpdfstring{$f$}{}}
\begin{figure}[t!]
    \begin{center}  
        \includegraphics[width=70mm]{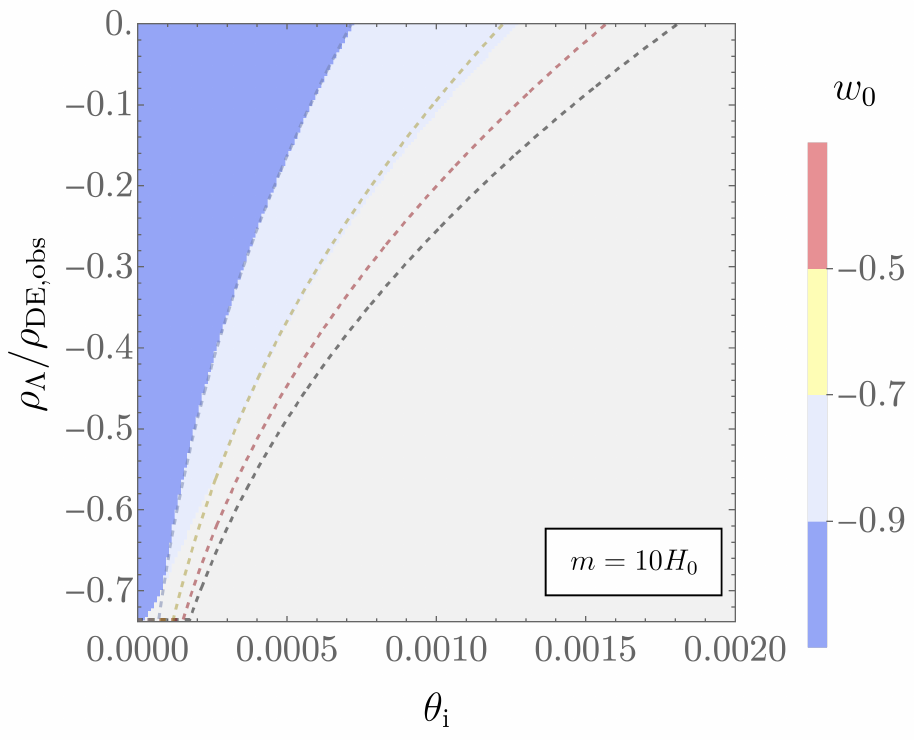}
        \\
        \vspace{2mm}
        \includegraphics[width=70mm]{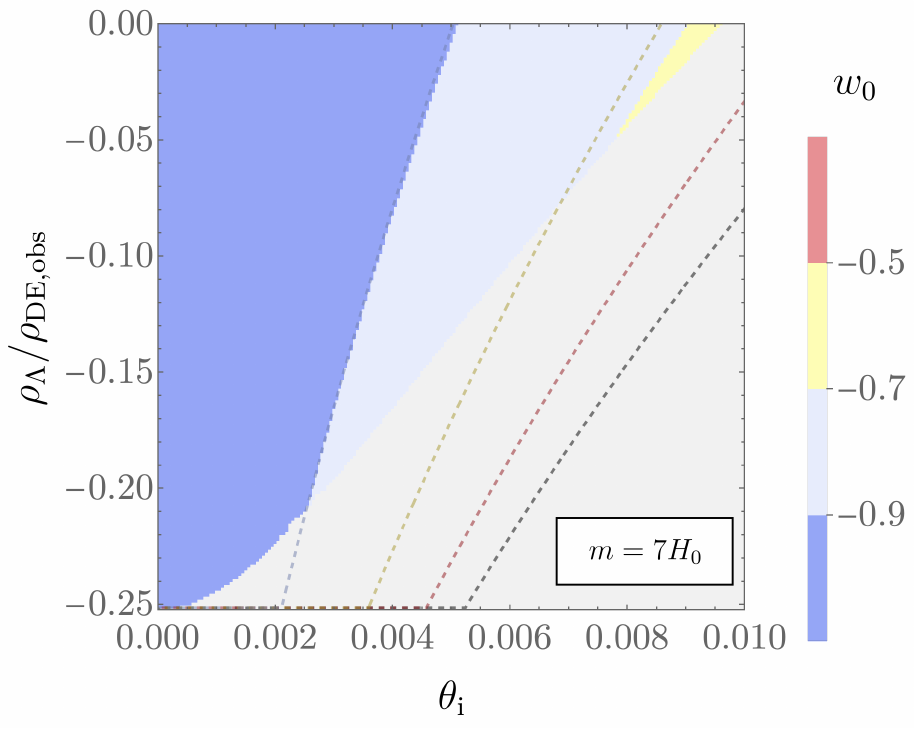}
        \\
        \vspace{2mm}
        \includegraphics[width=70mm]{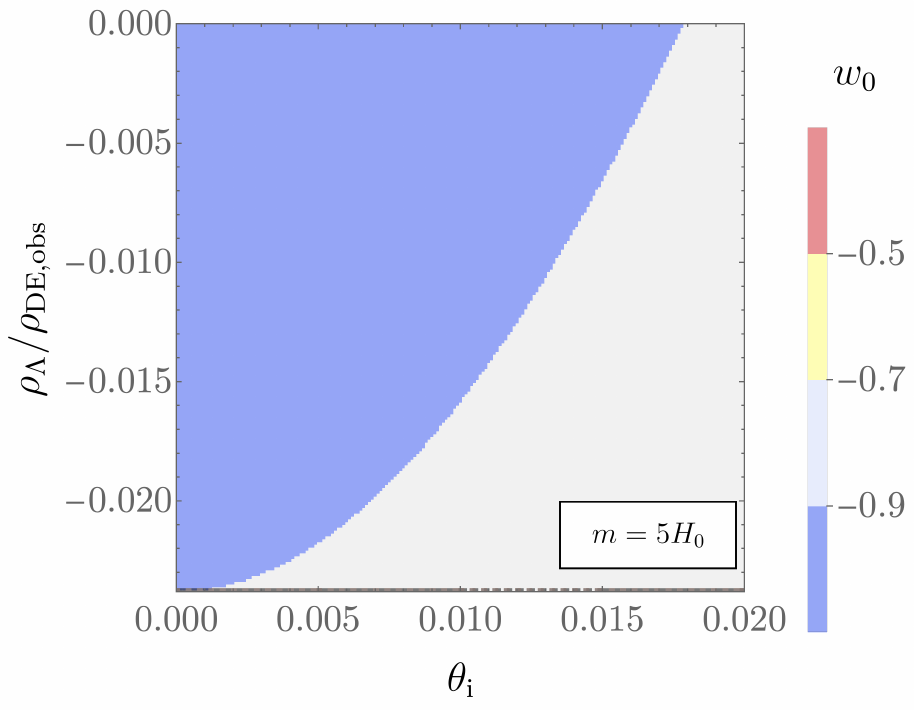}
        \end{center}
    \caption{%
        Same as Fig.~\ref{fig: w} except for the values of $m$ and $f = 0.1 M_\mathrm{Pl}$.
    }
    \label{fig: w smallf} 
\end{figure}

% \label{}
%%%%%%%%%%%%%%%%%%%%%%%%%%%%%%%%%%%%%%%
Next, we consider smaller decay constants.
We show the allowed parameter regions for $f = 0.1 M_\mathrm{Pl}$ with $m = 10H_0$, $7 H_0$, and $5 H_0$ in Fig.~\ref{fig: w smallf}.
For $m = 5 H_0$, $\rho_\Lambda$ should be near zero to realize the accelerated universe with the limited energy budget of the axion, $\rho_\phi \leq 2 m^2 f^2$.
Similarly to the bottom panel in Fig.~\ref{fig: w}, the allowed value of $w_0$ does not deviate largely from $-1$.
For $m = 7 H_0$ and $10H_0$, larger $w_0$ is allowed for large $\rho_\Lambda$.
%%%%%%%%%%%%%%%%%% Figure %%%%%%%%%%%%%%%%%%
%%%%%%%%%%%%%%%%%%%%%%%%%%%%%%%%%%%%%%%%%%%%

\begin{figure}[t!]
    \begin{center}  
        \includegraphics[width=80mm]{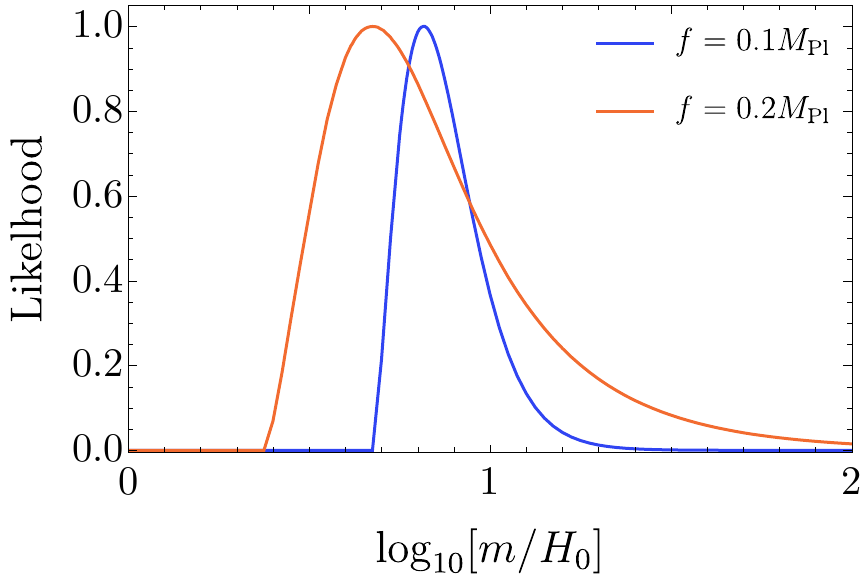}
        \end{center}
    \caption{%
        Same as Fig.~\ref{fig: mass likelihood} except for $f = 0.1 M_\mathrm{Pl}$ (blue) and $0.2 M_\mathrm{Pl}$ (red).
    }
    \label{fig: mass likelihood smallf} 
\end{figure}

We show the likelihood for $\ln m$ with $f = 0.1 M_\mathrm{Pl}$ and $0.2 M_\mathrm{Pl}$ in Fig.~\ref{fig: mass likelihood smallf}.
The mechanisms of the suppression for small and large masses are the same as in Fig.~\ref{fig: mass likelihood}.
However, the suppression for large mass is more severe for smaller $f$.
This is because the smaller Hubble parameter for the same $m$ requires finer tuning for $\theta_\mathrm{i}$.
As a result, the peak mass becomes smaller than the estimation by $2 m_\mathrm{peak}^2 f^2 = \rho_\Lambda^\mathrm{max}$.
%%%%%%%%%%%%%%%%%% Figure %%%%%%%%%%%%%%%%%%
%%%%%%%%%%%%%%%%%%%%%%%%%%%%%%%%%%%%%%%%%%%%

\begin{figure}[t!]
    \begin{center}  
        \includegraphics[width=80mm]{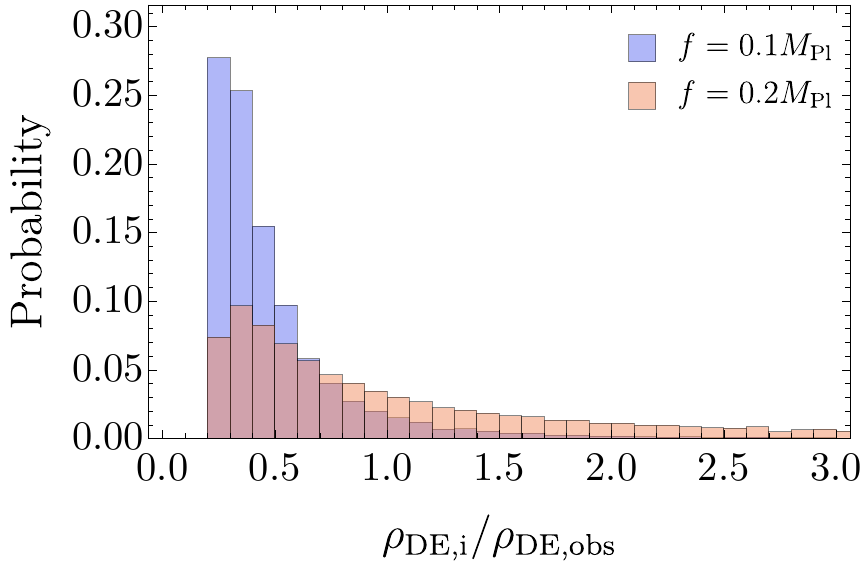}
        \end{center}
    \caption{%
       Same as Fig.~\ref{fig: rhoDE histogram} except for $f = 0.1 M_\mathrm{Pl}$ (blue) and $0.2 M_\mathrm{Pl}$ (red).
    }
    \label{fig: rhoDE histogram smallf} 
\end{figure}

Finally, we show the probability distribution for $\rho_\mathrm{DE,i}$ with $f = 0.1 M_\mathrm{Pl}$ and $0.2 M_\mathrm{Pl}$ in Fig.~\ref{fig: rhoDE histogram smallf}.
As mentioned above, the required tuning for $\theta_\mathrm{i}$ becomes more severe for smaller $f$.
As a result, $\rho_\mathrm{DE,i}$ becomes the order of the upper bound $\rho_\Lambda^\mathrm{max}$ only with a negligibly small probability.
Rather, $\rho_\mathrm{DE,i}$ typically lies around the observed value $\rho_\mathrm{DE, obs}$, which is the minimum order to realize the accelerated expansion.
While the probability distribution is concentrated below $\rho_{\rm DE,i} = \rho_{\rm DE,obs}$ for $f = 0.1 M_\mathrm{Pl}$, it spreads around $\rho_{\rm DE,i} \sim \rho_{\rm DE,obs}$ for $f = 0.2 M_\mathrm{Pl}$.
%%%%%%%%%%%%%%%%%% Figure %%%%%%%%%%%%%%%%%%
%%%%%%%%%%%%%%%%%%%%%%%%%%%%%%%%%%%%%%%%%%%%

%%%%%%%%%%%%%%% Section %%%%%%%%%%%%%%%
\section{Comparison with observational data}
\label{app: comparison}
%%%%%%%%%%%%%%%%%%%%%%%%%%%%%%%%%%%%%%%

Here, we compare our results with the latest observational analysis~\cite{Luu:2025dax}, which appeared on the arXiv after our submission.
The works of Refs.~\cite{Luu:2025fgw, Luu:2025dax} consider a dark energy model that consists of  of a cosmological constant and an axion and obtain the best-fit values of the model parameters from cosmological data including baryon acoustic oscillations (BAO), CMB, and Type Ia supernovae.
Focusing on a negative cosmological constant and an axion decay constant of $f = M_\mathrm{Pl}$, they find the best-fit values~\cite{Luu:2025dax}
\begin{equation}
\begin{gathered}
    \rho_\Lambda = -2.2 \rho_\mathrm{DE,obs}
    \ , \quad 
    m = 2.0 H_0 
    \ , 
    \\
    \theta_i = 0.18 \pi 
    \ , \quad 
    w_0 = -0.74
    \ ,
    \label{eq: bestfit}
\end{gathered}
\end{equation}
in our notation.
Since these values are obtained from observational data, the total dark energy density matches the observed value, which lies at the edge of the probability distribution in our anthropic discussion (Fig.~\ref{fig: rhoDE histogram}).
Correspondingly, the preferred axion mass also lies on the low-mass side of the peak for the likelihood of $\ln m$ shown in Fig.~\ref{fig: mass likelihood}.
The best-fit parameters are allowed by anthropic considerations, but they are not necessarily favored. For instance, in the case of the cosmological constant, the anthropic upper bound is about three orders of magnitude larger than the observed value. This may suggest that the anthropic conditions we have considered are not sufficient and that additional, more stringent conditions may exist, or it may simply be a coincidence. We note that, as discussed in the previous section, if the decay constant is slightly smaller than $M_{\rm Pl}$, the likelihood of the best-fit parameters can be enhanced.

We also compare our results with BAO.
The observation of BAO is typically quantified by the ratio of the comoving distance $D_M(z)$ to the redshift $z$ and the sound horizon $r_d$.
While $r_d$ is determined by the cosmic expansion before the recombination and is insensitive to the properties of dark energy, $D_M(z)$ is directly affected by the late-time expansion history.
When we take $\rho_\mathrm{DE,i} = \mathcal{O}(10^3) \rho_\mathrm{DE,obs}$ as a typical value in our anthropic discussion, the resultant value of $D_M(z)/r_d$ for fixed $z$ becomes much smaller than the value suggested by the observations, reflecting the larger Hubble parameter in the recent universe.

As another quantity representing the cosmological evolution, we can evaluate the lifetime of the universe in the presence of a negative cosmological constant.
Using the best-fit parameters~\eqref{eq: bestfit}, the total lifetime of the universe is estimated as $T \simeq 33.3$\,Gyr~\cite{Luu:2025dax}.
Here, we roughly estimate the lifetime of the universe for typical parameters in our anthropic discussion.
First, according to Eq.~\eqref{eq: phi approx}, the axion starts to oscillate at $t = t_\mathrm{osc} \sim m^{-1}$ up to logarithmic corrections.
Once the axion starts to oscillate and comes to behave as non-relativistic matter, the universe recollapses in a timescale of $\mathcal{O}(H_\mathrm{osc}^{-1})$ with $H_\mathrm{osc}$ being the Hubble parameter at $t = t_\mathrm{osc}$~\cite{Luu:2025dax}.
Thus, the lifetime of the universe is roughly estimated as
\begin{align}
    T 
    \sim 
    m^{-1}
    +
    \sqrt{\frac{\rho_\mathrm{DE,obs}}{\rho_\mathrm{DE,i}}} H_0^{-1} 
    \ ,
\end{align}
which is shorter than the age of the universe $\sim H_0^{-1}$ for typical parameters, $m = \mathcal{O}(10\,\text{--}\,100)H_0$ and $\rho_\mathrm{DE,i} = \mathcal{O}(10^3) \rho_\mathrm{DE,obs}$.
This implies that the universe often recollapses too early in our anthropic argument. Perhaps the emergence of intellectual life like ours requires a significantly longer cosmic lifetime, which may point to additional anthropic constraints beyond those considered here, or to a slightly smaller axion decay constant.

%%%%%%%%%%%% References %%%%%%%%%%%%%%
\bibliographystyle{apsrev4-1}
\bibliography{ref}

%merlin.mbs apsrev4-1.bst 2010-07-25 4.21a (PWD, AO, DPC) hacked
%Control: key (0)
%Control: author (72) initials jnrlst
%Control: editor formatted (1) identically to author
%Control: production of article title (-1) disabled
%Control: page (0) single
%Control: year (1) truncated
%Control: production of eprint (0) enabled
\begin{thebibliography}{54}%
\makeatletter
\providecommand \@ifxundefined [1]{%
 \@ifx{#1\undefined}
}%
\providecommand \@ifnum [1]{%
 \ifnum #1\expandafter \@firstoftwo
 \else \expandafter \@secondoftwo
 \fi
}%
\providecommand \@ifx [1]{%
 \ifx #1\expandafter \@firstoftwo
 \else \expandafter \@secondoftwo
 \fi
}%
\providecommand \natexlab [1]{#1}%
\providecommand \enquote  [1]{``#1''}%
\providecommand \bibnamefont  [1]{#1}%
\providecommand \bibfnamefont [1]{#1}%
\providecommand \citenamefont [1]{#1}%
\providecommand \href@noop [0]{\@secondoftwo}%
\providecommand \href [0]{\begingroup \@sanitize@url \@href}%
\providecommand \@href[1]{\@@startlink{#1}\@@href}%
\providecommand \@@href[1]{\endgroup#1\@@endlink}%
\providecommand \@sanitize@url [0]{\catcode `\\12\catcode `\$12\catcode
  `\&12\catcode `\#12\catcode `\^12\catcode `\_12\catcode `\%12\relax}%
\providecommand \@@startlink[1]{}%
\providecommand \@@endlink[0]{}%
\providecommand \url  [0]{\begingroup\@sanitize@url \@url }%
\providecommand \@url [1]{\endgroup\@href {#1}{\urlprefix }}%
\providecommand \urlprefix  [0]{URL }%
\providecommand \Eprint [0]{\href }%
\providecommand \doibase [0]{http://dx.doi.org/}%
\providecommand \selectlanguage [0]{\@gobble}%
\providecommand \bibinfo  [0]{\@secondoftwo}%
\providecommand \bibfield  [0]{\@secondoftwo}%
\providecommand \translation [1]{[#1]}%
\providecommand \BibitemOpen [0]{}%
\providecommand \bibitemStop [0]{}%
\providecommand \bibitemNoStop [0]{.\EOS\space}%
\providecommand \EOS [0]{\spacefactor3000\relax}%
\providecommand \BibitemShut  [1]{\csname bibitem#1\endcsname}%
\let\auto@bib@innerbib\@empty
%</preamble>
\bibitem [{\citenamefont {Perlmutter}\ \emph {et~al.}(1999)\citenamefont
  {Perlmutter} \emph {et~al.}}]{SupernovaCosmologyProject:1998vns}%
  \BibitemOpen
  \bibfield  {author} {\bibinfo {author} {\bibfnamefont {S.}~\bibnamefont
  {Perlmutter}} \emph {et~al.} (\bibinfo {collaboration} {Supernova Cosmology
  Project}),\ }\href {\doibase 10.1086/307221} {\bibfield  {journal} {\bibinfo
  {journal} {Astrophys. J.}\ }\textbf {\bibinfo {volume} {517}},\ \bibinfo
  {pages} {565} (\bibinfo {year} {1999})},\ \Eprint
  {http://arxiv.org/abs/astro-ph/9812133} {arXiv:astro-ph/9812133} \BibitemShut
  {NoStop}%
\bibitem [{\citenamefont {Riess}\ \emph {et~al.}(1998)\citenamefont {Riess}
  \emph {et~al.}}]{SupernovaSearchTeam:1998fmf}%
  \BibitemOpen
  \bibfield  {author} {\bibinfo {author} {\bibfnamefont {A.~G.}\ \bibnamefont
  {Riess}} \emph {et~al.} (\bibinfo {collaboration} {Supernova Search Team}),\
  }\href {\doibase 10.1086/300499} {\bibfield  {journal} {\bibinfo  {journal}
  {Astron. J.}\ }\textbf {\bibinfo {volume} {116}},\ \bibinfo {pages} {1009}
  (\bibinfo {year} {1998})},\ \Eprint {http://arxiv.org/abs/astro-ph/9805201}
  {arXiv:astro-ph/9805201} \BibitemShut {NoStop}%
\bibitem [{\citenamefont {Aghanim}\ \emph {et~al.}(2020)\citenamefont {Aghanim}
  \emph {et~al.}}]{Planck:2018vyg}%
  \BibitemOpen
  \bibfield  {author} {\bibinfo {author} {\bibfnamefont {N.}~\bibnamefont
  {Aghanim}} \emph {et~al.} (\bibinfo {collaboration} {Planck}),\ }\href
  {\doibase 10.1051/0004-6361/201833910} {\bibfield  {journal} {\bibinfo
  {journal} {Astron. Astrophys.}\ }\textbf {\bibinfo {volume} {641}},\ \bibinfo
  {pages} {A6} (\bibinfo {year} {2020})},\ \bibinfo {note} {[Erratum:
  Astron.Astrophys. 652, C4 (2021)]},\ \Eprint
  {http://arxiv.org/abs/1807.06209} {arXiv:1807.06209 [astro-ph.CO]}
  \BibitemShut {NoStop}%
\bibitem [{\citenamefont {Alam}\ \emph {et~al.}(2017)\citenamefont {Alam} \emph
  {et~al.}}]{BOSS:2016wmc}%
  \BibitemOpen
  \bibfield  {author} {\bibinfo {author} {\bibfnamefont {S.}~\bibnamefont
  {Alam}} \emph {et~al.} (\bibinfo {collaboration} {BOSS}),\ }\href {\doibase
  10.1093/mnras/stx721} {\bibfield  {journal} {\bibinfo  {journal} {Mon. Not.
  Roy. Astron. Soc.}\ }\textbf {\bibinfo {volume} {470}},\ \bibinfo {pages}
  {2617} (\bibinfo {year} {2017})},\ \Eprint {http://arxiv.org/abs/1607.03155}
  {arXiv:1607.03155 [astro-ph.CO]} \BibitemShut {NoStop}%
\bibitem [{\citenamefont {Carroll}\ \emph {et~al.}(2004)\citenamefont
  {Carroll}, \citenamefont {Duvvuri}, \citenamefont {Trodden},\ and\
  \citenamefont {Turner}}]{Carroll:2003wy}%
  \BibitemOpen
  \bibfield  {author} {\bibinfo {author} {\bibfnamefont {S.~M.}\ \bibnamefont
  {Carroll}}, \bibinfo {author} {\bibfnamefont {V.}~\bibnamefont {Duvvuri}},
  \bibinfo {author} {\bibfnamefont {M.}~\bibnamefont {Trodden}}, \ and\
  \bibinfo {author} {\bibfnamefont {M.~S.}\ \bibnamefont {Turner}},\ }\href
  {\doibase 10.1103/PhysRevD.70.043528} {\bibfield  {journal} {\bibinfo
  {journal} {Phys. Rev. D}\ }\textbf {\bibinfo {volume} {70}},\ \bibinfo
  {pages} {043528} (\bibinfo {year} {2004})},\ \Eprint
  {http://arxiv.org/abs/astro-ph/0306438} {arXiv:astro-ph/0306438} \BibitemShut
  {NoStop}%
\bibitem [{\citenamefont {Ratra}\ and\ \citenamefont
  {Peebles}(1988)}]{Ratra:1987rm}%
  \BibitemOpen
  \bibfield  {author} {\bibinfo {author} {\bibfnamefont {B.}~\bibnamefont
  {Ratra}}\ and\ \bibinfo {author} {\bibfnamefont {P.~J.~E.}\ \bibnamefont
  {Peebles}},\ }\href {\doibase 10.1103/PhysRevD.37.3406} {\bibfield  {journal}
  {\bibinfo  {journal} {Phys. Rev. D}\ }\textbf {\bibinfo {volume} {37}},\
  \bibinfo {pages} {3406} (\bibinfo {year} {1988})}\BibitemShut {NoStop}%
\bibitem [{\citenamefont {Caldwell}\ \emph {et~al.}(1998)\citenamefont
  {Caldwell}, \citenamefont {Dave},\ and\ \citenamefont
  {Steinhardt}}]{Caldwell:1997ii}%
  \BibitemOpen
  \bibfield  {author} {\bibinfo {author} {\bibfnamefont {R.~R.}\ \bibnamefont
  {Caldwell}}, \bibinfo {author} {\bibfnamefont {R.}~\bibnamefont {Dave}}, \
  and\ \bibinfo {author} {\bibfnamefont {P.~J.}\ \bibnamefont {Steinhardt}},\
  }\href {\doibase 10.1103/PhysRevLett.80.1582} {\bibfield  {journal} {\bibinfo
   {journal} {Phys. Rev. Lett.}\ }\textbf {\bibinfo {volume} {80}},\ \bibinfo
  {pages} {1582} (\bibinfo {year} {1998})},\ \Eprint
  {http://arxiv.org/abs/astro-ph/9708069} {arXiv:astro-ph/9708069} \BibitemShut
  {NoStop}%
\bibitem [{\citenamefont {Weinberg}(1989)}]{Weinberg:1988cp}%
  \BibitemOpen
  \bibfield  {author} {\bibinfo {author} {\bibfnamefont {S.}~\bibnamefont
  {Weinberg}},\ }\href {\doibase 10.1103/RevModPhys.61.1} {\bibfield  {journal}
  {\bibinfo  {journal} {Rev. Mod. Phys.}\ }\textbf {\bibinfo {volume} {61}},\
  \bibinfo {pages} {1} (\bibinfo {year} {1989})}\BibitemShut {NoStop}%
\bibitem [{\citenamefont {Kachru}\ \emph {et~al.}(2003)\citenamefont {Kachru},
  \citenamefont {Kallosh}, \citenamefont {Linde},\ and\ \citenamefont
  {Trivedi}}]{Kachru:2003aw}%
  \BibitemOpen
  \bibfield  {author} {\bibinfo {author} {\bibfnamefont {S.}~\bibnamefont
  {Kachru}}, \bibinfo {author} {\bibfnamefont {R.}~\bibnamefont {Kallosh}},
  \bibinfo {author} {\bibfnamefont {A.~D.}\ \bibnamefont {Linde}}, \ and\
  \bibinfo {author} {\bibfnamefont {S.~P.}\ \bibnamefont {Trivedi}},\ }\href
  {\doibase 10.1103/PhysRevD.68.046005} {\bibfield  {journal} {\bibinfo
  {journal} {Phys. Rev. D}\ }\textbf {\bibinfo {volume} {68}},\ \bibinfo
  {pages} {046005} (\bibinfo {year} {2003})},\ \Eprint
  {http://arxiv.org/abs/hep-th/0301240} {arXiv:hep-th/0301240} \BibitemShut
  {NoStop}%
\bibitem [{\citenamefont {Bousso}\ and\ \citenamefont
  {Polchinski}(2000)}]{Bousso:2000xa}%
  \BibitemOpen
  \bibfield  {author} {\bibinfo {author} {\bibfnamefont {R.}~\bibnamefont
  {Bousso}}\ and\ \bibinfo {author} {\bibfnamefont {J.}~\bibnamefont
  {Polchinski}},\ }\href {\doibase 10.1088/1126-6708/2000/06/006} {\bibfield
  {journal} {\bibinfo  {journal} {JHEP}\ }\textbf {\bibinfo {volume} {06}},\
  \bibinfo {pages} {006} (\bibinfo {year} {2000})},\ \Eprint
  {http://arxiv.org/abs/hep-th/0004134} {arXiv:hep-th/0004134} \BibitemShut
  {NoStop}%
\bibitem [{\citenamefont {Obied}\ \emph {et~al.}(2018)\citenamefont {Obied},
  \citenamefont {Ooguri}, \citenamefont {Spodyneiko},\ and\ \citenamefont
  {Vafa}}]{Obied:2018sgi}%
  \BibitemOpen
  \bibfield  {author} {\bibinfo {author} {\bibfnamefont {G.}~\bibnamefont
  {Obied}}, \bibinfo {author} {\bibfnamefont {H.}~\bibnamefont {Ooguri}},
  \bibinfo {author} {\bibfnamefont {L.}~\bibnamefont {Spodyneiko}}, \ and\
  \bibinfo {author} {\bibfnamefont {C.}~\bibnamefont {Vafa}},\ }\href@noop {}
  {\  (\bibinfo {year} {2018})},\ \Eprint {http://arxiv.org/abs/1806.08362}
  {arXiv:1806.08362 [hep-th]} \BibitemShut {NoStop}%
\bibitem [{\citenamefont {Danielsson}\ and\ \citenamefont
  {Van~Riet}(2018)}]{Danielsson:2018ztv}%
  \BibitemOpen
  \bibfield  {author} {\bibinfo {author} {\bibfnamefont {U.~H.}\ \bibnamefont
  {Danielsson}}\ and\ \bibinfo {author} {\bibfnamefont {T.}~\bibnamefont
  {Van~Riet}},\ }\href {\doibase 10.1142/S0218271818300070} {\bibfield
  {journal} {\bibinfo  {journal} {Int. J. Mod. Phys. D}\ }\textbf {\bibinfo
  {volume} {27}},\ \bibinfo {pages} {1830007} (\bibinfo {year} {2018})},\
  \Eprint {http://arxiv.org/abs/1804.01120} {arXiv:1804.01120 [hep-th]}
  \BibitemShut {NoStop}%
\bibitem [{\citenamefont {Demirtas}\ \emph {et~al.}(2022)\citenamefont
  {Demirtas}, \citenamefont {Kim}, \citenamefont {McAllister}, \citenamefont
  {Moritz},\ and\ \citenamefont {Rios-Tascon}}]{Demirtas:2021ote}%
  \BibitemOpen
  \bibfield  {author} {\bibinfo {author} {\bibfnamefont {M.}~\bibnamefont
  {Demirtas}}, \bibinfo {author} {\bibfnamefont {M.}~\bibnamefont {Kim}},
  \bibinfo {author} {\bibfnamefont {L.}~\bibnamefont {McAllister}}, \bibinfo
  {author} {\bibfnamefont {J.}~\bibnamefont {Moritz}}, \ and\ \bibinfo {author}
  {\bibfnamefont {A.}~\bibnamefont {Rios-Tascon}},\ }\href {\doibase
  10.1103/PhysRevLett.128.011602} {\bibfield  {journal} {\bibinfo  {journal}
  {Phys. Rev. Lett.}\ }\textbf {\bibinfo {volume} {128}},\ \bibinfo {pages}
  {011602} (\bibinfo {year} {2022})},\ \Eprint
  {http://arxiv.org/abs/2107.09065} {arXiv:2107.09065 [hep-th]} \BibitemShut
  {NoStop}%
\bibitem [{\citenamefont {Weinberg}(1987)}]{Weinberg:1987dv}%
  \BibitemOpen
  \bibfield  {author} {\bibinfo {author} {\bibfnamefont {S.}~\bibnamefont
  {Weinberg}},\ }\href {\doibase 10.1103/PhysRevLett.59.2607} {\bibfield
  {journal} {\bibinfo  {journal} {Phys. Rev. Lett.}\ }\textbf {\bibinfo
  {volume} {59}},\ \bibinfo {pages} {2607} (\bibinfo {year}
  {1987})}\BibitemShut {NoStop}%
\bibitem [{\citenamefont {Susskind}(2003)}]{Susskind:2003kw}%
  \BibitemOpen
  \bibfield  {author} {\bibinfo {author} {\bibfnamefont {L.}~\bibnamefont
  {Susskind}},\ }\href@noop {} {\ ,\ \bibinfo {pages} {247} (\bibinfo {year}
  {2003})},\ \Eprint {http://arxiv.org/abs/hep-th/0302219}
  {arXiv:hep-th/0302219} \BibitemShut {NoStop}%
\bibitem [{\citenamefont {Polchinski}(2006)}]{Polchinski:2006gy}%
  \BibitemOpen
  \bibfield  {author} {\bibinfo {author} {\bibfnamefont {J.}~\bibnamefont
  {Polchinski}},\ }in\ \href@noop {} {\emph {\bibinfo {booktitle} {{23rd Solvay
  Conference in Physics: The Quantum Structure of Space and Time}}}}\ (\bibinfo
  {year} {2006})\ pp.\ \bibinfo {pages} {216--236},\ \Eprint
  {http://arxiv.org/abs/hep-th/0603249} {arXiv:hep-th/0603249} \BibitemShut
  {NoStop}%
\bibitem [{\citenamefont {Ellis}\ and\ \citenamefont
  {Smolin}(2009)}]{Ellis:2009gx}%
  \BibitemOpen
  \bibfield  {author} {\bibinfo {author} {\bibfnamefont {G.~F.~R.}\
  \bibnamefont {Ellis}}\ and\ \bibinfo {author} {\bibfnamefont
  {L.}~\bibnamefont {Smolin}},\ }\href@noop {} {\  (\bibinfo {year} {2009})},\
  \Eprint {http://arxiv.org/abs/0901.2414} {arXiv:0901.2414 [hep-th]}
  \BibitemShut {NoStop}%
\bibitem [{\citenamefont {Sorini}\ \emph {et~al.}(2024)\citenamefont {Sorini},
  \citenamefont {Peacock},\ and\ \citenamefont {Lombriser}}]{Sorini:2024tto}%
  \BibitemOpen
  \bibfield  {author} {\bibinfo {author} {\bibfnamefont {D.}~\bibnamefont
  {Sorini}}, \bibinfo {author} {\bibfnamefont {J.~A.}\ \bibnamefont {Peacock}},
  \ and\ \bibinfo {author} {\bibfnamefont {L.}~\bibnamefont {Lombriser}},\
  }\href {\doibase 10.1093/mnras/stae2236} {\bibfield  {journal} {\bibinfo
  {journal} {Mon. Not. Roy. Astron. Soc.}\ }\textbf {\bibinfo {volume} {535}},\
  \bibinfo {pages} {1449} (\bibinfo {year} {2024})},\ \Eprint
  {http://arxiv.org/abs/2411.07301} {arXiv:2411.07301 [astro-ph.CO]}
  \BibitemShut {NoStop}%
\bibitem [{\citenamefont {Takahashi}(2009)}]{Takahashi:2008pu}%
  \BibitemOpen
  \bibfield  {author} {\bibinfo {author} {\bibfnamefont {F.}~\bibnamefont
  {Takahashi}},\ }\href {\doibase 10.1143/PTP.121.711} {\bibfield  {journal}
  {\bibinfo  {journal} {Prog. Theor. Phys.}\ }\textbf {\bibinfo {volume}
  {121}},\ \bibinfo {pages} {711} (\bibinfo {year} {2009})},\ \Eprint
  {http://arxiv.org/abs/0804.2478} {arXiv:0804.2478 [hep-ph]} \BibitemShut
  {NoStop}%
\bibitem [{\citenamefont {Kaloper}(2019)}]{Kaloper:2018kma}%
  \BibitemOpen
  \bibfield  {author} {\bibinfo {author} {\bibfnamefont {N.}~\bibnamefont
  {Kaloper}},\ }\href {\doibase 10.1007/JHEP11(2019)106} {\bibfield  {journal}
  {\bibinfo  {journal} {JHEP}\ }\textbf {\bibinfo {volume} {11}},\ \bibinfo
  {pages} {106} (\bibinfo {year} {2019})},\ \Eprint
  {http://arxiv.org/abs/1806.03308} {arXiv:1806.03308 [hep-th]} \BibitemShut
  {NoStop}%
\bibitem [{\citenamefont {Arvanitaki}\ \emph {et~al.}(2010)\citenamefont
  {Arvanitaki}, \citenamefont {Dimopoulos}, \citenamefont {Dubovsky},
  \citenamefont {Kaloper},\ and\ \citenamefont
  {March-Russell}}]{Arvanitaki:2009fg}%
  \BibitemOpen
  \bibfield  {author} {\bibinfo {author} {\bibfnamefont {A.}~\bibnamefont
  {Arvanitaki}}, \bibinfo {author} {\bibfnamefont {S.}~\bibnamefont
  {Dimopoulos}}, \bibinfo {author} {\bibfnamefont {S.}~\bibnamefont
  {Dubovsky}}, \bibinfo {author} {\bibfnamefont {N.}~\bibnamefont {Kaloper}}, \
  and\ \bibinfo {author} {\bibfnamefont {J.}~\bibnamefont {March-Russell}},\
  }\href {\doibase 10.1103/PhysRevD.81.123530} {\bibfield  {journal} {\bibinfo
  {journal} {Phys. Rev. D}\ }\textbf {\bibinfo {volume} {81}},\ \bibinfo
  {pages} {123530} (\bibinfo {year} {2010})},\ \Eprint
  {http://arxiv.org/abs/0905.4720} {arXiv:0905.4720 [hep-th]} \BibitemShut
  {NoStop}%
\bibitem [{\citenamefont {Svrcek}(2006)}]{Svrcek:2006hf}%
  \BibitemOpen
  \bibfield  {author} {\bibinfo {author} {\bibfnamefont {P.}~\bibnamefont
  {Svrcek}},\ }\href@noop {} {\  (\bibinfo {year} {2006})},\ \Eprint
  {http://arxiv.org/abs/hep-th/0607086} {arXiv:hep-th/0607086} \BibitemShut
  {NoStop}%
\bibitem [{\citenamefont {Ruchika}\ \emph {et~al.}(2023)\citenamefont
  {Ruchika}, \citenamefont {Adil}, \citenamefont {Dutta}, \citenamefont
  {Mukherjee},\ and\ \citenamefont {Sen}}]{Ruchika:2020avj}%
  \BibitemOpen
  \bibfield  {author} {\bibinfo {author} {\bibnamefont {Ruchika}}, \bibinfo
  {author} {\bibfnamefont {S.~A.}\ \bibnamefont {Adil}}, \bibinfo {author}
  {\bibfnamefont {K.}~\bibnamefont {Dutta}}, \bibinfo {author} {\bibfnamefont
  {A.}~\bibnamefont {Mukherjee}}, \ and\ \bibinfo {author} {\bibfnamefont
  {A.~A.}\ \bibnamefont {Sen}},\ }\href {\doibase 10.1016/j.dark.2023.101199}
  {\bibfield  {journal} {\bibinfo  {journal} {Phys. Dark Univ.}\ }\textbf
  {\bibinfo {volume} {40}},\ \bibinfo {pages} {101199} (\bibinfo {year}
  {2023})},\ \Eprint {http://arxiv.org/abs/2005.08813} {arXiv:2005.08813
  [astro-ph.CO]} \BibitemShut {NoStop}%
\bibitem [{\citenamefont {Luu}\ \emph {et~al.}(2025{\natexlab{a}})\citenamefont
  {Luu}, \citenamefont {Qiu},\ and\ \citenamefont {Tye}}]{Luu:2025fgw}%
  \BibitemOpen
  \bibfield  {author} {\bibinfo {author} {\bibfnamefont {H.~N.}\ \bibnamefont
  {Luu}}, \bibinfo {author} {\bibfnamefont {Y.-C.}\ \bibnamefont {Qiu}}, \ and\
  \bibinfo {author} {\bibfnamefont {S.~H.~H.}\ \bibnamefont {Tye}},\
  }\href@noop {} {\  (\bibinfo {year} {2025}{\natexlab{a}})},\ \Eprint
  {http://arxiv.org/abs/2503.18120} {arXiv:2503.18120 [hep-ph]} \BibitemShut
  {NoStop}%
\bibitem [{\citenamefont {Cardenas}\ \emph {et~al.}(2003)\citenamefont
  {Cardenas}, \citenamefont {Gonzalez}, \citenamefont {Leiva}, \citenamefont
  {Martin},\ and\ \citenamefont {Quiros}}]{Cardenas:2002np}%
  \BibitemOpen
  \bibfield  {author} {\bibinfo {author} {\bibfnamefont {R.}~\bibnamefont
  {Cardenas}}, \bibinfo {author} {\bibfnamefont {T.}~\bibnamefont {Gonzalez}},
  \bibinfo {author} {\bibfnamefont {Y.}~\bibnamefont {Leiva}}, \bibinfo
  {author} {\bibfnamefont {O.}~\bibnamefont {Martin}}, \ and\ \bibinfo {author}
  {\bibfnamefont {I.}~\bibnamefont {Quiros}},\ }\href {\doibase
  10.1103/PhysRevD.67.083501} {\bibfield  {journal} {\bibinfo  {journal} {Phys.
  Rev. D}\ }\textbf {\bibinfo {volume} {67}},\ \bibinfo {pages} {083501}
  (\bibinfo {year} {2003})},\ \Eprint {http://arxiv.org/abs/astro-ph/0206315}
  {arXiv:astro-ph/0206315} \BibitemShut {NoStop}%
\bibitem [{\citenamefont {Dutta}\ \emph {et~al.}(2020)\citenamefont {Dutta},
  \citenamefont {Ruchika}, \citenamefont {Roy}, \citenamefont {Sen},\ and\
  \citenamefont {Sheikh-Jabbari}}]{Dutta:2018vmq}%
  \BibitemOpen
  \bibfield  {author} {\bibinfo {author} {\bibfnamefont {K.}~\bibnamefont
  {Dutta}}, \bibinfo {author} {\bibnamefont {Ruchika}}, \bibinfo {author}
  {\bibfnamefont {A.}~\bibnamefont {Roy}}, \bibinfo {author} {\bibfnamefont
  {A.~A.}\ \bibnamefont {Sen}}, \ and\ \bibinfo {author} {\bibfnamefont
  {M.~M.}\ \bibnamefont {Sheikh-Jabbari}},\ }\href {\doibase
  10.1007/s10714-020-2665-4} {\bibfield  {journal} {\bibinfo  {journal} {Gen.
  Rel. Grav.}\ }\textbf {\bibinfo {volume} {52}},\ \bibinfo {pages} {15}
  (\bibinfo {year} {2020})},\ \Eprint {http://arxiv.org/abs/1808.06623}
  {arXiv:1808.06623 [astro-ph.CO]} \BibitemShut {NoStop}%
\bibitem [{\citenamefont {Visinelli}\ \emph {et~al.}(2019)\citenamefont
  {Visinelli}, \citenamefont {Vagnozzi},\ and\ \citenamefont
  {Danielsson}}]{Visinelli:2019qqu}%
  \BibitemOpen
  \bibfield  {author} {\bibinfo {author} {\bibfnamefont {L.}~\bibnamefont
  {Visinelli}}, \bibinfo {author} {\bibfnamefont {S.}~\bibnamefont {Vagnozzi}},
  \ and\ \bibinfo {author} {\bibfnamefont {U.}~\bibnamefont {Danielsson}},\
  }\href {\doibase 10.3390/sym11081035} {\bibfield  {journal} {\bibinfo
  {journal} {Symmetry}\ }\textbf {\bibinfo {volume} {11}},\ \bibinfo {pages}
  {1035} (\bibinfo {year} {2019})},\ \Eprint {http://arxiv.org/abs/1907.07953}
  {arXiv:1907.07953 [astro-ph.CO]} \BibitemShut {NoStop}%
\bibitem [{\citenamefont {Calder\'on}\ \emph {et~al.}(2021)\citenamefont
  {Calder\'on}, \citenamefont {Gannouji}, \citenamefont {L'Huillier},\ and\
  \citenamefont {Polarski}}]{Calderon:2020hoc}%
  \BibitemOpen
  \bibfield  {author} {\bibinfo {author} {\bibfnamefont {R.}~\bibnamefont
  {Calder\'on}}, \bibinfo {author} {\bibfnamefont {R.}~\bibnamefont
  {Gannouji}}, \bibinfo {author} {\bibfnamefont {B.}~\bibnamefont
  {L'Huillier}}, \ and\ \bibinfo {author} {\bibfnamefont {D.}~\bibnamefont
  {Polarski}},\ }\href {\doibase 10.1103/PhysRevD.103.023526} {\bibfield
  {journal} {\bibinfo  {journal} {Phys. Rev. D}\ }\textbf {\bibinfo {volume}
  {103}},\ \bibinfo {pages} {023526} (\bibinfo {year} {2021})},\ \Eprint
  {http://arxiv.org/abs/2008.10237} {arXiv:2008.10237 [astro-ph.CO]}
  \BibitemShut {NoStop}%
\bibitem [{\citenamefont {Sen}\ \emph {et~al.}(2022)\citenamefont {Sen},
  \citenamefont {Adil},\ and\ \citenamefont {Sen}}]{Sen:2021wld}%
  \BibitemOpen
  \bibfield  {author} {\bibinfo {author} {\bibfnamefont {A.~A.}\ \bibnamefont
  {Sen}}, \bibinfo {author} {\bibfnamefont {S.~A.}\ \bibnamefont {Adil}}, \
  and\ \bibinfo {author} {\bibfnamefont {S.}~\bibnamefont {Sen}},\ }\href
  {\doibase 10.1093/mnras/stac2796} {\bibfield  {journal} {\bibinfo  {journal}
  {Mon. Not. Roy. Astron. Soc.}\ }\textbf {\bibinfo {volume} {518}},\ \bibinfo
  {pages} {1098} (\bibinfo {year} {2022})},\ \Eprint
  {http://arxiv.org/abs/2112.10641} {arXiv:2112.10641 [astro-ph.CO]}
  \BibitemShut {NoStop}%
\bibitem [{\citenamefont {Adil}\ \emph {et~al.}(2023)\citenamefont {Adil},
  \citenamefont {Mukhopadhyay}, \citenamefont {Sen},\ and\ \citenamefont
  {Vagnozzi}}]{Adil:2023ara}%
  \BibitemOpen
  \bibfield  {author} {\bibinfo {author} {\bibfnamefont {S.~A.}\ \bibnamefont
  {Adil}}, \bibinfo {author} {\bibfnamefont {U.}~\bibnamefont {Mukhopadhyay}},
  \bibinfo {author} {\bibfnamefont {A.~A.}\ \bibnamefont {Sen}}, \ and\
  \bibinfo {author} {\bibfnamefont {S.}~\bibnamefont {Vagnozzi}},\ }\href
  {\doibase 10.1088/1475-7516/2023/10/072} {\bibfield  {journal} {\bibinfo
  {journal} {JCAP}\ }\textbf {\bibinfo {volume} {10}},\ \bibinfo {pages} {072}
  (\bibinfo {year} {2023})},\ \Eprint {http://arxiv.org/abs/2307.12763}
  {arXiv:2307.12763 [astro-ph.CO]} \BibitemShut {NoStop}%
\bibitem [{\citenamefont {Menci}\ \emph {et~al.}(2024)\citenamefont {Menci},
  \citenamefont {Adil}, \citenamefont {Mukhopadhyay}, \citenamefont {Sen},\
  and\ \citenamefont {Vagnozzi}}]{Menci:2024rbq}%
  \BibitemOpen
  \bibfield  {author} {\bibinfo {author} {\bibfnamefont {N.}~\bibnamefont
  {Menci}}, \bibinfo {author} {\bibfnamefont {S.~A.}\ \bibnamefont {Adil}},
  \bibinfo {author} {\bibfnamefont {U.}~\bibnamefont {Mukhopadhyay}}, \bibinfo
  {author} {\bibfnamefont {A.~A.}\ \bibnamefont {Sen}}, \ and\ \bibinfo
  {author} {\bibfnamefont {S.}~\bibnamefont {Vagnozzi}},\ }\href {\doibase
  10.1088/1475-7516/2024/07/072} {\bibfield  {journal} {\bibinfo  {journal}
  {JCAP}\ }\textbf {\bibinfo {volume} {07}},\ \bibinfo {pages} {072} (\bibinfo
  {year} {2024})},\ \Eprint {http://arxiv.org/abs/2401.12659} {arXiv:2401.12659
  [astro-ph.CO]} \BibitemShut {NoStop}%
\bibitem [{\citenamefont {Adame}\ \emph {et~al.}(2025)\citenamefont {Adame}
  \emph {et~al.}}]{DESI:2024mwx}%
  \BibitemOpen
  \bibfield  {author} {\bibinfo {author} {\bibfnamefont {A.~G.}\ \bibnamefont
  {Adame}} \emph {et~al.} (\bibinfo {collaboration} {DESI}),\ }\href {\doibase
  10.1088/1475-7516/2025/02/021} {\bibfield  {journal} {\bibinfo  {journal}
  {JCAP}\ }\textbf {\bibinfo {volume} {02}},\ \bibinfo {pages} {021} (\bibinfo
  {year} {2025})},\ \Eprint {http://arxiv.org/abs/2404.03002} {arXiv:2404.03002
  [astro-ph.CO]} \BibitemShut {NoStop}%
\bibitem [{\citenamefont {Abdul~Karim}\ \emph
  {et~al.}(2025{\natexlab{a}})\citenamefont {Abdul~Karim} \emph
  {et~al.}}]{DESI:2025zgx}%
  \BibitemOpen
  \bibfield  {author} {\bibinfo {author} {\bibfnamefont {M.}~\bibnamefont
  {Abdul~Karim}} \emph {et~al.} (\bibinfo {collaboration} {DESI}),\ }\href@noop
  {} {\  (\bibinfo {year} {2025}{\natexlab{a}})},\ \Eprint
  {http://arxiv.org/abs/2503.14738} {arXiv:2503.14738 [astro-ph.CO]}
  \BibitemShut {NoStop}%
\bibitem [{\citenamefont {Abdul~Karim}\ \emph
  {et~al.}(2025{\natexlab{b}})\citenamefont {Abdul~Karim} \emph
  {et~al.}}]{DESI:2025zpo}%
  \BibitemOpen
  \bibfield  {author} {\bibinfo {author} {\bibfnamefont {M.}~\bibnamefont
  {Abdul~Karim}} \emph {et~al.} (\bibinfo {collaboration} {DESI}),\ }\href@noop
  {} {\  (\bibinfo {year} {2025}{\natexlab{b}})},\ \Eprint
  {http://arxiv.org/abs/2503.14739} {arXiv:2503.14739 [astro-ph.CO]}
  \BibitemShut {NoStop}%
\bibitem [{\citenamefont {Tada}\ and\ \citenamefont
  {Terada}(2024)}]{Tada:2024znt}%
  \BibitemOpen
  \bibfield  {author} {\bibinfo {author} {\bibfnamefont {Y.}~\bibnamefont
  {Tada}}\ and\ \bibinfo {author} {\bibfnamefont {T.}~\bibnamefont {Terada}},\
  }\href {\doibase 10.1103/PhysRevD.109.L121305} {\bibfield  {journal}
  {\bibinfo  {journal} {Phys. Rev. D}\ }\textbf {\bibinfo {volume} {109}},\
  \bibinfo {pages} {L121305} (\bibinfo {year} {2024})},\ \Eprint
  {http://arxiv.org/abs/2404.05722} {arXiv:2404.05722 [astro-ph.CO]}
  \BibitemShut {NoStop}%
\bibitem [{\citenamefont {Yin}(2024)}]{Yin:2024hba}%
  \BibitemOpen
  \bibfield  {author} {\bibinfo {author} {\bibfnamefont {W.}~\bibnamefont
  {Yin}},\ }\href {\doibase 10.1007/JHEP05(2024)327} {\bibfield  {journal}
  {\bibinfo  {journal} {JHEP}\ }\textbf {\bibinfo {volume} {05}},\ \bibinfo
  {pages} {327} (\bibinfo {year} {2024})},\ \Eprint
  {http://arxiv.org/abs/2404.06444} {arXiv:2404.06444 [hep-ph]} \BibitemShut
  {NoStop}%
\bibitem [{\citenamefont {Cort\^es}\ and\ \citenamefont
  {Liddle}(2024)}]{Cortes:2024lgw}%
  \BibitemOpen
  \bibfield  {author} {\bibinfo {author} {\bibfnamefont {M.}~\bibnamefont
  {Cort\^es}}\ and\ \bibinfo {author} {\bibfnamefont {A.~R.}\ \bibnamefont
  {Liddle}},\ }\href {\doibase 10.1088/1475-7516/2024/12/007} {\bibfield
  {journal} {\bibinfo  {journal} {JCAP}\ }\textbf {\bibinfo {volume} {12}},\
  \bibinfo {pages} {007} (\bibinfo {year} {2024})},\ \Eprint
  {http://arxiv.org/abs/2404.08056} {arXiv:2404.08056 [astro-ph.CO]}
  \BibitemShut {NoStop}%
\bibitem [{\citenamefont {Bhattacharya}\ \emph
  {et~al.}(2024{\natexlab{a}})\citenamefont {Bhattacharya}, \citenamefont
  {Borghetto}, \citenamefont {Malhotra}, \citenamefont {Parameswaran},
  \citenamefont {Tasinato},\ and\ \citenamefont
  {Zavala}}]{Bhattacharya:2024hep}%
  \BibitemOpen
  \bibfield  {author} {\bibinfo {author} {\bibfnamefont {S.}~\bibnamefont
  {Bhattacharya}}, \bibinfo {author} {\bibfnamefont {G.}~\bibnamefont
  {Borghetto}}, \bibinfo {author} {\bibfnamefont {A.}~\bibnamefont {Malhotra}},
  \bibinfo {author} {\bibfnamefont {S.}~\bibnamefont {Parameswaran}}, \bibinfo
  {author} {\bibfnamefont {G.}~\bibnamefont {Tasinato}}, \ and\ \bibinfo
  {author} {\bibfnamefont {I.}~\bibnamefont {Zavala}},\ }\href {\doibase
  10.1088/1475-7516/2024/09/073} {\bibfield  {journal} {\bibinfo  {journal}
  {JCAP}\ }\textbf {\bibinfo {volume} {09}},\ \bibinfo {pages} {073} (\bibinfo
  {year} {2024}{\natexlab{a}})},\ \Eprint {http://arxiv.org/abs/2405.17396}
  {arXiv:2405.17396 [astro-ph.CO]} \BibitemShut {NoStop}%
\bibitem [{\citenamefont {Mukherjee}\ and\ \citenamefont
  {Sen}(2024)}]{Mukherjee:2024ryz}%
  \BibitemOpen
  \bibfield  {author} {\bibinfo {author} {\bibfnamefont {P.}~\bibnamefont
  {Mukherjee}}\ and\ \bibinfo {author} {\bibfnamefont {A.~A.}\ \bibnamefont
  {Sen}},\ }\href {\doibase 10.1103/PhysRevD.110.123502} {\bibfield  {journal}
  {\bibinfo  {journal} {Phys. Rev. D}\ }\textbf {\bibinfo {volume} {110}},\
  \bibinfo {pages} {123502} (\bibinfo {year} {2024})},\ \Eprint
  {http://arxiv.org/abs/2405.19178} {arXiv:2405.19178 [astro-ph.CO]}
  \BibitemShut {NoStop}%
\bibitem [{\citenamefont {Notari}\ \emph {et~al.}(2024)\citenamefont {Notari},
  \citenamefont {Redi},\ and\ \citenamefont {Tesi}}]{Notari:2024rti}%
  \BibitemOpen
  \bibfield  {author} {\bibinfo {author} {\bibfnamefont {A.}~\bibnamefont
  {Notari}}, \bibinfo {author} {\bibfnamefont {M.}~\bibnamefont {Redi}}, \ and\
  \bibinfo {author} {\bibfnamefont {A.}~\bibnamefont {Tesi}},\ }\href {\doibase
  10.1088/1475-7516/2024/11/025} {\bibfield  {journal} {\bibinfo  {journal}
  {JCAP}\ }\textbf {\bibinfo {volume} {11}},\ \bibinfo {pages} {025} (\bibinfo
  {year} {2024})},\ \Eprint {http://arxiv.org/abs/2406.08459} {arXiv:2406.08459
  [astro-ph.CO]} \BibitemShut {NoStop}%
\bibitem [{\citenamefont {Jia}\ \emph {et~al.}(2025)\citenamefont {Jia},
  \citenamefont {Hu}, \citenamefont {Yi},\ and\ \citenamefont
  {Wang}}]{Jia:2024wix}%
  \BibitemOpen
  \bibfield  {author} {\bibinfo {author} {\bibfnamefont {X.~D.}\ \bibnamefont
  {Jia}}, \bibinfo {author} {\bibfnamefont {J.~P.}\ \bibnamefont {Hu}},
  \bibinfo {author} {\bibfnamefont {S.~X.}\ \bibnamefont {Yi}}, \ and\ \bibinfo
  {author} {\bibfnamefont {F.~Y.}\ \bibnamefont {Wang}},\ }\href {\doibase
  10.3847/2041-8213/ada94d} {\bibfield  {journal} {\bibinfo  {journal}
  {Astrophys. J. Lett.}\ }\textbf {\bibinfo {volume} {979}},\ \bibinfo {pages}
  {L34} (\bibinfo {year} {2025})},\ \Eprint {http://arxiv.org/abs/2406.02019}
  {arXiv:2406.02019 [astro-ph.CO]} \BibitemShut {NoStop}%
\bibitem [{\citenamefont {Hern\'andez-Almada}\ \emph
  {et~al.}(2024)\citenamefont {Hern\'andez-Almada}, \citenamefont
  {Mendoza-Mart\'\i{}nez}, \citenamefont {Garc\'\i{}a-Aspeitia},\ and\
  \citenamefont {Motta}}]{Hernandez-Almada:2024ost}%
  \BibitemOpen
  \bibfield  {author} {\bibinfo {author} {\bibfnamefont {A.}~\bibnamefont
  {Hern\'andez-Almada}}, \bibinfo {author} {\bibfnamefont {M.~L.}\ \bibnamefont
  {Mendoza-Mart\'\i{}nez}}, \bibinfo {author} {\bibfnamefont {M.~A.}\
  \bibnamefont {Garc\'\i{}a-Aspeitia}}, \ and\ \bibinfo {author} {\bibfnamefont
  {V.}~\bibnamefont {Motta}},\ }\href {\doibase 10.1016/j.dark.2024.101668}
  {\bibfield  {journal} {\bibinfo  {journal} {Phys. Dark Univ.}\ }\textbf
  {\bibinfo {volume} {46}},\ \bibinfo {pages} {101668} (\bibinfo {year}
  {2024})},\ \Eprint {http://arxiv.org/abs/2407.09430} {arXiv:2407.09430
  [astro-ph.CO]} \BibitemShut {NoStop}%
\bibitem [{\citenamefont {Bhattacharya}\ \emph
  {et~al.}(2024{\natexlab{b}})\citenamefont {Bhattacharya}, \citenamefont
  {Borghetto}, \citenamefont {Malhotra}, \citenamefont {Parameswaran},
  \citenamefont {Tasinato},\ and\ \citenamefont
  {Zavala}}]{Bhattacharya:2024kxp}%
  \BibitemOpen
  \bibfield  {author} {\bibinfo {author} {\bibfnamefont {S.}~\bibnamefont
  {Bhattacharya}}, \bibinfo {author} {\bibfnamefont {G.}~\bibnamefont
  {Borghetto}}, \bibinfo {author} {\bibfnamefont {A.}~\bibnamefont {Malhotra}},
  \bibinfo {author} {\bibfnamefont {S.}~\bibnamefont {Parameswaran}}, \bibinfo
  {author} {\bibfnamefont {G.}~\bibnamefont {Tasinato}}, \ and\ \bibinfo
  {author} {\bibfnamefont {I.}~\bibnamefont {Zavala}},\ }\href@noop {} {\
  (\bibinfo {year} {2024}{\natexlab{b}})},\ \Eprint
  {http://arxiv.org/abs/2410.21243} {arXiv:2410.21243 [astro-ph.CO]}
  \BibitemShut {NoStop}%
\bibitem [{\citenamefont {Berbig}(2025)}]{Berbig:2024aee}%
  \BibitemOpen
  \bibfield  {author} {\bibinfo {author} {\bibfnamefont {M.}~\bibnamefont
  {Berbig}},\ }\href {\doibase 10.1088/1475-7516/2025/03/015} {\bibfield
  {journal} {\bibinfo  {journal} {JCAP}\ }\textbf {\bibinfo {volume} {03}},\
  \bibinfo {pages} {015} (\bibinfo {year} {2025})},\ \Eprint
  {http://arxiv.org/abs/2412.07418} {arXiv:2412.07418 [astro-ph.CO]}
  \BibitemShut {NoStop}%
\bibitem [{\citenamefont {Wolf}\ \emph {et~al.}(2025)\citenamefont {Wolf},
  \citenamefont {Garc\'\i{}a-Garc\'\i{}a},\ and\ \citenamefont
  {Ferreira}}]{Wolf:2025jlc}%
  \BibitemOpen
  \bibfield  {author} {\bibinfo {author} {\bibfnamefont {W.~J.}\ \bibnamefont
  {Wolf}}, \bibinfo {author} {\bibfnamefont {C.}~\bibnamefont
  {Garc\'\i{}a-Garc\'\i{}a}}, \ and\ \bibinfo {author} {\bibfnamefont {P.~G.}\
  \bibnamefont {Ferreira}},\ }\href@noop {} {\  (\bibinfo {year} {2025})},\
  \Eprint {http://arxiv.org/abs/2502.04929} {arXiv:2502.04929 [astro-ph.CO]}
  \BibitemShut {NoStop}%
\bibitem [{\citenamefont {Chakraborty}\ \emph {et~al.}(2025)\citenamefont
  {Chakraborty}, \citenamefont {Chanda}, \citenamefont {Das},\ and\
  \citenamefont {Dutta}}]{Chakraborty:2025syu}%
  \BibitemOpen
  \bibfield  {author} {\bibinfo {author} {\bibfnamefont {A.}~\bibnamefont
  {Chakraborty}}, \bibinfo {author} {\bibfnamefont {P.~K.}\ \bibnamefont
  {Chanda}}, \bibinfo {author} {\bibfnamefont {S.}~\bibnamefont {Das}}, \ and\
  \bibinfo {author} {\bibfnamefont {K.}~\bibnamefont {Dutta}},\ }\href@noop {}
  {\  (\bibinfo {year} {2025})},\ \Eprint {http://arxiv.org/abs/2503.10806}
  {arXiv:2503.10806 [astro-ph.CO]} \BibitemShut {NoStop}%
\bibitem [{\citenamefont {Borghetto}\ \emph {et~al.}(2025)\citenamefont
  {Borghetto}, \citenamefont {Malhotra}, \citenamefont {Tasinato},\ and\
  \citenamefont {Zavala}}]{Borghetto:2025jrk}%
  \BibitemOpen
  \bibfield  {author} {\bibinfo {author} {\bibfnamefont {G.}~\bibnamefont
  {Borghetto}}, \bibinfo {author} {\bibfnamefont {A.}~\bibnamefont {Malhotra}},
  \bibinfo {author} {\bibfnamefont {G.}~\bibnamefont {Tasinato}}, \ and\
  \bibinfo {author} {\bibfnamefont {I.}~\bibnamefont {Zavala}},\ }\href@noop {}
  {\  (\bibinfo {year} {2025})},\ \Eprint {http://arxiv.org/abs/2503.11628}
  {arXiv:2503.11628 [astro-ph.CO]} \BibitemShut {NoStop}%
\bibitem [{\citenamefont {Nakagawa}\ \emph {et~al.}(2025)\citenamefont
  {Nakagawa}, \citenamefont {Nakai}, \citenamefont {Qiu},\ and\ \citenamefont
  {Yamada}}]{Nakagawa:2025ejs}%
  \BibitemOpen
  \bibfield  {author} {\bibinfo {author} {\bibfnamefont {S.}~\bibnamefont
  {Nakagawa}}, \bibinfo {author} {\bibfnamefont {Y.}~\bibnamefont {Nakai}},
  \bibinfo {author} {\bibfnamefont {Y.-C.}\ \bibnamefont {Qiu}}, \ and\
  \bibinfo {author} {\bibfnamefont {M.}~\bibnamefont {Yamada}},\ }\href@noop {}
  {\  (\bibinfo {year} {2025})},\ \Eprint {http://arxiv.org/abs/2503.18924}
  {arXiv:2503.18924 [astro-ph.CO]} \BibitemShut {NoStop}%
\bibitem [{\citenamefont {Lee}\ \emph {et~al.}(2025)\citenamefont {Lee},
  \citenamefont {Murai}, \citenamefont {Takahashi},\ and\ \citenamefont
  {Yin}}]{Lee:2025yvn}%
  \BibitemOpen
  \bibfield  {author} {\bibinfo {author} {\bibfnamefont {J.}~\bibnamefont
  {Lee}}, \bibinfo {author} {\bibfnamefont {K.}~\bibnamefont {Murai}}, \bibinfo
  {author} {\bibfnamefont {F.}~\bibnamefont {Takahashi}}, \ and\ \bibinfo
  {author} {\bibfnamefont {W.}~\bibnamefont {Yin}},\ }\href@noop {} {\
  (\bibinfo {year} {2025})},\ \Eprint {http://arxiv.org/abs/2503.18417}
  {arXiv:2503.18417 [hep-ph]} \BibitemShut {NoStop}%
\bibitem [{\citenamefont {Colg\'ain}\ \emph {et~al.}(2025)\citenamefont
  {Colg\'ain}, \citenamefont {Pourojaghi}, \citenamefont {Sheikh-Jabbari},\
  and\ \citenamefont {Yin}}]{Colgain:2025nzf}%
  \BibitemOpen
  \bibfield  {author} {\bibinfo {author} {\bibfnamefont {E.~O.}\ \bibnamefont
  {Colg\'ain}}, \bibinfo {author} {\bibfnamefont {S.}~\bibnamefont
  {Pourojaghi}}, \bibinfo {author} {\bibfnamefont {M.~M.}\ \bibnamefont
  {Sheikh-Jabbari}}, \ and\ \bibinfo {author} {\bibfnamefont {L.}~\bibnamefont
  {Yin}},\ }\href@noop {} {\  (\bibinfo {year} {2025})},\ \Eprint
  {http://arxiv.org/abs/2504.04417} {arXiv:2504.04417 [astro-ph.CO]}
  \BibitemShut {NoStop}%
\bibitem [{\citenamefont {Santos}\ \emph {et~al.}(2025)\citenamefont {Santos},
  \citenamefont {Morais}, \citenamefont {Pan}, \citenamefont {Yang},\ and\
  \citenamefont {Di~Valentino}}]{Santos:2025wiv}%
  \BibitemOpen
  \bibfield  {author} {\bibinfo {author} {\bibfnamefont {F.~B. M.~d.}\
  \bibnamefont {Santos}}, \bibinfo {author} {\bibfnamefont {J.}~\bibnamefont
  {Morais}}, \bibinfo {author} {\bibfnamefont {S.}~\bibnamefont {Pan}},
  \bibinfo {author} {\bibfnamefont {W.}~\bibnamefont {Yang}}, \ and\ \bibinfo
  {author} {\bibfnamefont {E.}~\bibnamefont {Di~Valentino}},\ }\href@noop {} {\
   (\bibinfo {year} {2025})},\ \Eprint {http://arxiv.org/abs/2504.04646}
  {arXiv:2504.04646 [astro-ph.CO]} \BibitemShut {NoStop}%
\bibitem [{\citenamefont {Hellerman}\ and\ \citenamefont
  {Walcher}(2005)}]{Hellerman:2005yi}%
  \BibitemOpen
  \bibfield  {author} {\bibinfo {author} {\bibfnamefont {S.}~\bibnamefont
  {Hellerman}}\ and\ \bibinfo {author} {\bibfnamefont {J.}~\bibnamefont
  {Walcher}},\ }\href {\doibase 10.1103/PhysRevD.72.123520} {\bibfield
  {journal} {\bibinfo  {journal} {Phys. Rev. D}\ }\textbf {\bibinfo {volume}
  {72}},\ \bibinfo {pages} {123520} (\bibinfo {year} {2005})},\ \Eprint
  {http://arxiv.org/abs/hep-th/0508161} {arXiv:hep-th/0508161} \BibitemShut
  {NoStop}%
\bibitem [{\citenamefont {Takahashi}\ and\ \citenamefont
  {Yamada}(2019)}]{Takahashi:2019ypv}%
  \BibitemOpen
  \bibfield  {author} {\bibinfo {author} {\bibfnamefont {F.}~\bibnamefont
  {Takahashi}}\ and\ \bibinfo {author} {\bibfnamefont {M.}~\bibnamefont
  {Yamada}},\ }\href {\doibase 10.1088/1475-7516/2019/07/001} {\bibfield
  {journal} {\bibinfo  {journal} {JCAP}\ }\textbf {\bibinfo {volume} {07}},\
  \bibinfo {pages} {001} (\bibinfo {year} {2019})},\ \Eprint
  {http://arxiv.org/abs/1904.12864} {arXiv:1904.12864 [astro-ph.CO]}
  \BibitemShut {NoStop}%
\bibitem [{\citenamefont {Luu}\ \emph {et~al.}(2025{\natexlab{b}})\citenamefont
  {Luu}, \citenamefont {Qiu},\ and\ \citenamefont {Tye}}]{Luu:2025dax}%
  \BibitemOpen
  \bibfield  {author} {\bibinfo {author} {\bibfnamefont {H.~N.}\ \bibnamefont
  {Luu}}, \bibinfo {author} {\bibfnamefont {Y.-C.}\ \bibnamefont {Qiu}}, \ and\
  \bibinfo {author} {\bibfnamefont {S.~H.~H.}\ \bibnamefont {Tye}},\
  }\href@noop {} {\  (\bibinfo {year} {2025}{\natexlab{b}})},\ \Eprint
  {http://arxiv.org/abs/2506.24011} {arXiv:2506.24011 [hep-ph]} \BibitemShut
  {NoStop}%
\end{thebibliography}%
%%%%%%%%%%%%%%%%%%%%%%%%%%%%%%%%%%%%%%

\end{document}